\documentclass[aps,twocolumn,prx,showpacs,superscriptaddress,10pt,floatfix]{revtex4-2}
\usepackage{graphicx}
\usepackage[normalem]{ulem}
\usepackage{amsmath,nicefrac}
\usepackage{amssymb}
\usepackage{xcolor}
\usepackage{braket}
\usepackage{enumerate}
\usepackage{ifthen}
\usepackage{xspace}
\usepackage[caption=false]{subfig}
\usepackage{physics}
\usepackage{braket}
\usepackage{dcolumn}% Align table columns on decimal point
\usepackage{bm}% bold math
\usepackage{hyperref}
\usepackage{ytableau}
\usepackage{comment}
\usepackage{lipsum}
\usepackage{mathrsfs}

\newcommand{\beq}{\begin{equation}} 							
\newcommand{\eeq}{\end{equation}}
\newcommand{\bematrix}{\left(\begin{matrix}}
\newcommand{\ematrix}{\end{matrix}\right)}
\newcommand{\id}{\mathbb{I}}
\newcommand{\zf}{Z_\mathrm{F}}
\newcommand{\zb}{Z_\mathrm{B}}
\newcommand{\pf}{p_\mathrm{F}}

\def\cH{\mathcal H}

\def\cL{\mathcal L}

\def\cO{\mathcal O}

\def\cV{\mathcal V}

\def\C{{\ensuremath{\mathbb C}}}

\def\R{{\ensuremath{\mathbb R}}}

\def\tr{\mathrm{Tr}}

\def\one{{\mbox{$1 \hspace{-1.0mm}  {\bf l}$}}}

\newcommand{\eqnref}[1]{Eq. (\ref{#1})}

\newcommand{\figref}[1]{Fig. \ref{#1}}

\newcommand{\secref}[1]{Sec. \ref{#1}}
\newcommand{\appref}[1]{App. \ref{#1}}

\newcommand{\citeref}[1]{Ref. \cite{#1}}

\newcommand{\changes}[1]{{\color{black} #1}}

\newtheorem{theorem}{Theorem}

\newtheorem{proposition}{Proposition}

\newtheorem{definition}{Definition}



\def\pf{p_{\textrm{F}}}
\def\UB{U_{\textrm{B}}}
\def\UF{U_{\textrm{F}}}

\def\pfX{\dfrac{\partial p_{\textrm{F}}}{\partial X}}
\def\pfXX{\dfrac{\partial^2 p_{\textrm{F}}}{\partial X^2}}

\def\phiX{\dfrac{\partial\phi}{\partial X}}

\def\Uomega{\dfrac{\partial U}{\partial \omega}}
\def\Uoomega{\dfrac{\partial^2 U}{\partial \omega^2}}

\def\Unu{\dfrac{\partial U}{\partial \nu}}
\def\Unnu{\dfrac{\partial^2 U}{\partial \nu^2}}

\def\UX{\dfrac{\partial U}{\partial X}}
\def\UXX{\dfrac{\partial^2 U}{\partial X^2}}

\def\UBX{\dfrac{\partial U_\textrm{B}}{\partial X}}
\def\UBXX{\dfrac{\partial^2 U_\textrm{B}}{\partial X^2}}

\def\UFX{\dfrac{\partial U_\textrm{F}}{\partial X}}
\def\UFXX{\dfrac{\partial^2 U_\textrm{F}}{\partial X^2}}

\def\UBomega{\dfrac{\partial U_\textrm{B}}{\partial \omega}}

\def\UBnu{\dfrac{\partial U_\textrm{B}}{\partial \nu}}

\def\UBbeta{\dfrac{\partial U_\textrm{B}}{\partial \beta}}
\def\UBbbeta{\dfrac{\partial^2 U_\textrm{B}}{\partial \beta^2}}

\def\UFomega{\dfrac{\partial U_\textrm{F}}{\partial \omega}}

\def\UFnu{\dfrac{\partial U_\textrm{F}}{\partial \nu}}

\def\UFbeta{\dfrac{\partial U_\textrm{F}}{\partial \beta}}

\def\nuX{\dfrac{\partial \nu}{\partial X}}

\begin{document}
\title{Thermodynamics of Hamiltonian anyons with applications to quantum heat engines}

\author{Joe Dunlop}\email[]{j.dunlop@exeter.ac.uk}
\affiliation{Physics and Astronomy, University of Exeter, Exeter EX4 4QL, United Kingdom\looseness=-1}

\author{\'Alvaro Tejero}\email[ ]{atejero@onsager.ugr.es}\affiliation{Electromagnetism and Condensed Matter Department, Universidad de Granada, 18071 Granada, Spain\looseness=-1}%
\affiliation{Instituto Carlos I de Física Teórica y Computacional, Universidad de Granada, 18071 Granada, Spain\looseness=-1}%

\author{Michalis Skotiniotis}\email[]{mskotiniotis@onsager.ugr.es}
\affiliation{Electromagnetism and Condensed Matter Department, Universidad de Granada, 18071 Granada, Spain\looseness=-1}%
\affiliation{Instituto Carlos I de Física Teórica y Computacional, Universidad de Granada, 18071 Granada, Spain\looseness=-1}%

\author{Daniel Manzano}\email[]{manzano@onsager.ugr.es}
\affiliation{Electromagnetism and Condensed Matter Department, Universidad de Granada, 18071 Granada, Spain\looseness=-1}%
\affiliation{Instituto Carlos I de Física Teórica y Computacional, Universidad de Granada, 18071 Granada, Spain\looseness=-1}%

\date{\today}
\begin{abstract}
    The behavior of a collection of identical particles is intimately linked to the symmetries of their wavefunction under particle exchange. Topological anyons, arising as quasiparticles in low-dimensional systems, interpolate between bosons and fermions, picking up a complex phase when exchanged. Recent research has demonstrated that similar statistical behavior can arise with mixtures of bosonic and fermionic pairs, offering theoretical and experimental simplicity. We introduce an alternative implementation of such \emph{statistical anyons}, based on promoting or suppressing the population of symmetric states via a symmetry generating Hamiltonian. The scheme has numerous advantages: anyonic statistics emerge in a single particle pair, extending straightforwardly to larger systems; the statistical properties can be dynamically adjusted; and the setup can be simulated efficiently. We show how exchange symmetry can be exploited to improve the performance of heat engines, and demonstrate a reversible work extraction cycle in which bosonization and fermionization replace compression and expansion strokes. Additionally, we investigate emergent thermal properties, including critical phenomena, in large statistical anyon systems.
\end{abstract}
\maketitle
\section{Introduction}

Symmetry under particle exchange is an essential property of quantum systems 
\cite{galindo_90,peres_95,bertlmann2023modern}. The indistinguishability of quantum particles means that 
swapping them results in no measurable difference. Exchanging a pair of identical particles in states 
$\ket{\psi_1}$ and $\ket{\psi_2}$ can therefore only change the joint state by a global phase
\begin{equation}
    \ket{\psi_2,\psi_1} = e^{i\pi\varphi}\ket{\psi_1,\psi_2},
\end{equation}
where $\varphi=2n$ for bosons and $\varphi=2n+1$ for fermions, with $n\in \mathbb{N}_0$. In three 
dimensions these are the only two possibilities due to the spin-statistics theorem \cite{pauli40}. 
However, quasiparticles in two dimensional systems can acquire an arbitrary phase. These particles, 
which do not conform to either fermionic or bosonic behavior, are known as anyons 
\cite{Wilczek_82,Wilczek_06} and play a pivotal role in the quantum Hall effect 
\cite{stern08,carrega_21} as well as universal topological quantum computing 
\cite{kitaev_03,lahtinen_17}. Experimental evidence for anyons has been observed 
via Aharonov-Bohm interference \cite{nakamura_20} and anyonic collisions \cite{bartolomei_20}. Recently, 
anyons have also been simulated on a trapped-ion processor \cite{iqbal_24}.

Exchange symmetry is deeply linked to the thermodynamic properties of identical particle systems. The 
antisymmetry of fermions implies that no two can share exactly the same state (the Pauli exclusion 
principle) leading to a repulsive exchange interaction; whereas bosons can cluster without restriction. This distinction underlies the difference between Fermi-Dirac and Bose-Einstein statistics  \cite{bose_24,einstein_24,dirac_26,fermi_26,landau_88}. Anyon systems, in contrast, exhibit fractional  statistics which can be modeled by a generalized exclusion principle  \cite{haldane_91} allowing their thermal properties to interpolate between those of fermions and bosons \cite{wu_94,murthy_94}.

\begin{figure}[t]
\centering
    \includegraphics[width=\columnwidth]{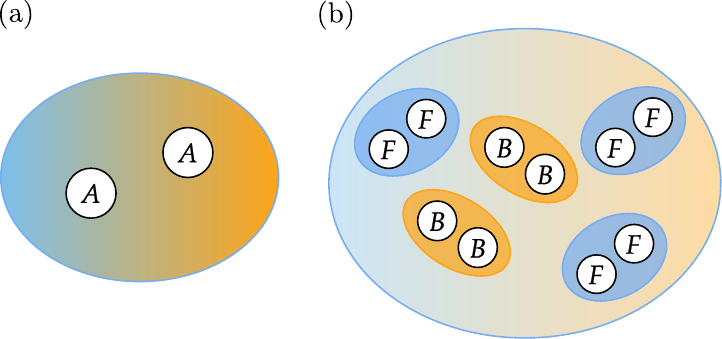}
    \caption{An illustration of two statistical approaches to anyons. Orange represents fermionic exchange symmetry, while blue indicates bosonic.
    (a) Hamiltonian anyons as introduced in this paper. A single pair of particles occupies a mixed quantum state which overlaps with both the symmetric and antisymmetric subspaces. The framework can easily be extended to $N>2$ particles.
    (b) Statistical anyons, introduced in \cite{myers_21} as an ensemble of particle pairs, each in a purely symmetric or antisymmetric state. Collectively, the statistical properties of the ensemble interpolate between those of bosons and fermions.
    }

\label{fig:anyons}
\end{figure}

The physical implementation of anyons is challenging. In Ref.~\cite{myers_21}, Myers and Deffner 
establish an alternative approach. Instead of analyzing the behavior of real low-dimensional 
\emph{topological} anyons, the authors consider an ensemble composed of pairs of particles with either 
bosonic or fermionic symmetry introducing the concept of \emph{statistical} anyons. This model is found 
to be equivalent to generalized exclusion statistics, which closely approximates the thermodynamic 
properties of topological anyons. In this paper, we present a novel approach to statistical anyons 
illustrated in Fig. \ref{fig:anyons}. Rather than working with an ensemble of bosonic and fermionic 
systems, we  instead consider a \emph{single} multi-particle system in a mixed quantum state---which 
overlaps with both symmetric and antisymmetric subspaces---with the intrinsic symmetry of the particles 
encoded into a latent degree of freedom. For a carefully engineered Hamiltonian the thermal equilibrium 
state interpolates between fermionic and bosonic symmetries. We refer to this new construction as 
\emph{Hamiltonian anyons}.

Hamiltonian anyons have a particularly interesting property: the balance between 
fermionic and bosonic statistics can be tuned, or even dynamically altered, via externally controlled 
Hamiltonian parameters or by changing the temperature. We find that, for larger systems, the crossover 
between bosonic and fermionic regimes results in first and second-order phase transitions which can be exploited 
for the construction of quantum heat engines.

Aside from their transformative technological applications, the study of heat engines has been pivotal in the conceptual development of thermodynamics, a tradition which continues in the modern field of quantum thermodynamics \cite{gemmer_09,kosloff13,vinjanampathy_16,binder_18,strasberg2022quantum}. Since the pioneering contributions of Alicki \cite{alicki_79}, Linden \cite{linden_10} and Scully \cite{scully_11}, many different models of quantum engine have been proposed, extending and refining the concepts of heat and work \cite{alicki_79,tejero_24}. This research also brings to light the thermodynamic role of uniquely quantum resources such as coherence and squeezing \cite{klaers_17,tejero_preprint}. 
\changes{In recent years, the role of particle statistics has been an active topic, with investigations into the comparative performance of fermions, bosons and anyons as a working medium \cite{myers_20,myers_21,mani_25}; enhancements from quantum phase transitions \cite{Myers_22,Eglinton_23}, and exotic statistical effects such as $q$-deformation \cite{ozaydin_23}.}
Quantum engines operating with interacting particles have also been studied \cite{Jaramillo2016,Chen2019,Menon_2025}, 
\changes{exploiting the ability to mimic Pauli exclusion and the resulting statistical properties of fermions via the repulsion of hard-core bosons.}
Building on this, a work extraction cycle has been experimentally demonstrated in which fermionization and bosonization, mediated by Feshbach resonance, replace the roles of heating and cooling \cite{koch_23}.

We here investigate two heat engines utilizing the unique advantages of Hamiltonian anyons. In the first, a Stirling cycle, the system is 
actively driven between fermionic and bosonic statistics, achieving Carnot efficiency in the 
limit of quasistatic and isothermal driving. In the second, an Otto cycle, we vary the frequency 
of the confining potential as a rapid quench, with the system transitioning passively between fermionic and bosonic 
regimes during heating and cooling. We find that the power of the cycle is 
substantially enhanced as compared to a purely fermionic or bosonic medium.

The paper is organized as follows. Section~\ref{sec:II} introduces the requisite mathematical background 
for the treatment of exchange symmetry in quantum mechanical systems. Section~\ref{sec:thermodynamics} 
introduces a model Hamiltonian and analyzes the equilibrium thermal properties of a multi-particle 
system before characterizing the fermion-boson phase transition. Based on this, two heat engines are 
proposed in Section~\ref{sec:engines}. In Section~\ref{sec:simulation} we consider the simulation of 
Hamiltonian anyons in digital quantum computers. We summarize and conclude in 
Section~\ref{sec:conclusions}.

\section{Hilbert spaces under exchange symmetry}
\label{sec:II}

This section explores the central idea behind the construction of multi-particle systems with 
varying statistical properties. The approach utilizes auxiliary degrees of freedom for each 
particle to control their collective statistical behavior under particle exchange. To keep the 
focus on conceptual understanding, we offer an expository description here, with the 
comprehensive mathematical treatment provided in Appendix \ref{app:hilbert}.

To illustrate the construction, consider a pair of confined particles, each with two degrees 
of freedom. For later clarity, we assume one degree of freedom relates to the energy of the 
particles, while the other corresponds to their spin. The state space for each particle is the 
tensor product
	\begin{equation}\label{eqn:statespace_1p}
		\mathcal{H} = \mathscr{H} \otimes \mathbb{C}^d,
	\end{equation}
where $\mathscr{H}$ is the countably infinite dimensional Hilbert space corresponding to the 
energy of the particle, and $\C^d$ is the $d$-dimensional space associated with its spin. For 
a spin-$j$ particle, the spin space dimension is $d = 2j + 1$. An orthonormal basis for 
$\mathscr{H}$ is
\begin{equation}\label{eqn:eigenbasis_fock}
    B_{\mathscr{H}} = \biggl\{\ket{n}\Big| \; n\in\mathbb{N}_0\biggr\} ,
    \end{equation}
where $\{ \ket{n} \}$ represents energy eigenstates. For a spin-$1/2$ particle, an orthonormal 
basis for $\mathbb{C}^2$ is
\begin{equation}\label{eqn:eigenbasis_2spins}
  B_{\C^2} =\{\ket{0}:=\ket{\uparrow}, \ket{1}:=\ket{\downarrow}\}, 
\end{equation}
so the total Hilbert space for a single particle, Eq. (\ref{eqn:statespace_1p}), has an 
orthonormal basis
\begin{equation}\label{eqn:eigenbasis_1p}
    B_{\cH} = \biggl\{\ket{n}\otimes\ket{m}\Big| \; n\in\mathbb{N}_0,  m\in\{0,1\}\biggr\}.
\end{equation}	
Consequently, the total state space for two \emph{distinguishable} spin-$1/2$ particles is 
$\mathcal{H}^{\otimes 2} = \mathscr{H}^{\otimes 2} \otimes (\mathbb{C}^2)^{\otimes 2}$, with 
an orthonormal basis
\begin{equation}\label{eqn:eigenbasis_2p}
\begin{split}
    & B_{\cH^{\otimes 2}} = \\ &\left\{\ket{n_1,n_2}\otimes \ket{m_1,m_2}\Big| \; n_i\in\mathbb{N}_0,  m_i\in\{0,1\}, i=1,2 \right\}.
    \end{split}
\end{equation} 
If the particles are \emph{indistinguishable} fermions their collective state, $\ket{\Psi}$, 
must be antisymmetric under particle exchange, i.e., it belongs to the totally antisymmetric 
subspace $\ket{\Psi}\in\cH^{(\mathrm{Alt})}\subset\cH^{\otimes 2}$. With respect to the spaces 
$\mathscr{H}^{\otimes 2}$ and $(\C^2 )^{\otimes 2}$, the antisymmetric subspace decomposes 
into a direct sum in the following fashion
	\begin{equation}\label{eqn:hilbert_alt_2p}
		\begin{split}
		 \cH^{(\mathrm{Alt})}&=\left(\left[\mathscr{H}^{\otimes 2}\right]^{(\text{Sym})}\otimes \left[\left(\C^2 \right)^{\otimes 2}\right]^{(\text{Alt})}\right)\\
		&\oplus \left( \left[\mathscr{H}^{\otimes 2}\right]^{(\text{Alt})}\otimes \left[\left(\C^2 \right)^{\otimes 2}\right]^{(\text{Sym})}\right)  ,
		\end{split}
	\end{equation}	   	
where 
	\begin{equation}\label{eqn:C2_sym}
     		\begin{split}
			\left[\left(\C^2 \right)^{\otimes 2}\right]^{(\text{Sym})} &= \textrm{span}\left\{ \ket{00},  \ket{01}+\ket{10},  \ket{11}\right\},\\
			\left[\left(\C^2 \right)^{\otimes 2}\right]^{(\text{Alt})} &=\textrm{span}\left\{\ket{01}-\ket{10}\right\},
		\end{split}
	\end{equation}  
are the familiar triplet and singlet subspaces respectively, and
	\begin{equation}\label{eqn:states_fock_sym}
		\begin{split}
			&\left[\mathscr{H}^{\otimes 2}\right]^{(\text{Sym})}   = \\
                &\hspace{3em}\textrm{span}\left\{\ket{n{-}k, k}+\ket{k, n{-}k},  0\leq k\leq \left\lfloor\frac{n}{2}\right\rfloor\right\},\\[1em]
			&\left[\mathscr{H}^{\otimes 2}\right]^{(\text{Alt})}   = \\
                &\hspace{3em}\textrm{span}\left\{\ket{n{-}k,k}-\ket{k,n{-}k}, 0\leq k\leq \left\lceil\frac{n}{2}\right\rceil{-}1\right\},
		\end{split}
	\end{equation}
are the symmetric and antisymmetric spaces corresponding to the energy eigenbasis. Note that 
we have used the shorthand $\ket{k,n-k}:=\ket{k}\otimes\ket{n-k}$ for two-particle states of 
total energy $n$.  

Thus, the energetic and spin degrees of freedom of two spin-$1/2$ fermions combine such that 
if the energetic degrees of freedom are symmetric under particle exchange, then the spin 
degrees of freedom are antisymmetric, and \emph{vice versa}.  Note that a similar argument 
would apply to bosons. In this case, their collective state belongs to the totally symmetric 
subspace $\cH^{(\mathrm{Sym})}\subset\cH^{\otimes 2}$, such that the energy and spin degrees 
of freedom are either both symmetric or antisymmetric under particle exchange. 

Suppose we trace over one degree of freedom of the two-particle system. The state for the 
remaining degree of freedom would then be an incoherent mixture of states that are either 
totally symmetric or antisymmetric, i.e., for $\ket{\Psi}\in\cH^{(\mathrm{Alt})}$
	\begin{equation}\label{eqn:stat_anyon}
		\rho=\tr_{\C^2}\ketbra{\Psi}{\Psi}=p \rho^{(\mathrm{Sym})} + (1-p) \rho^{(\mathrm{Alt})},
	\end{equation}
where $\tr_{\C^2}$ indicates that we have traced over the spin degrees of freedom, and 
$\rho^{(\mathrm{Sym})}\in\mathscr{B}\{[\mathscr{H}^{\otimes 2}]^{(\mathrm{Sym})}\}$, 
$\rho^{(\mathrm{Alt})}\in\mathscr{B}\{[\mathscr{H}^{\otimes 2}]^{(\mathrm{Alt})}\}$. Here, 
$\mathscr{B}(\cH)$ indicates the space of bounded linear operators acting on a Hilbert space 
$\cH$ with finite Hilbert-Schmidt norm (see \cite{conway07,takhtadzhian2008quantum} for 
further details). Thus, by carefully engineering the joint spin state of two particles, we can 
control the probability, $p$, with which their energy degrees of freedom are symmetric (resp. 
antisymmetric) under particle exchange. This results in a continuous interpolation between 
bosonic and fermionic statistics giving rise to so-called statistical anyons \cite{myers_21}. 

The above scenario can be generalized to the case of $N$ particles.  The total state space 
$\cH^{\otimes N}=\mathscr{H}^{\otimes N}\otimes (\C^d)^{\otimes N}$ can be decomposed into 
subspaces according to the various symmetries of the collective state of the particles under 
exchange, see \appref{app:hilbert}. If the particles are bosons (resp. fermions) then the 
totally symmetric (resp. antisymmetric) subspaces block-diagonalize as 
	\begin{equation}\label{eqn:hilbert_sym}
		\begin{split}
        		\cH^{(\text{Sym})}&=\left( \left[\mathscr{H}^{\otimes N}\right]^{(\text{Sym})}\otimes \left[\left(\C^d \right)^{\otimes N}\right]^{(\text{Sym})}\right) \\
        		&\oplus \left(\left[\mathscr{H}^{\otimes N}\right]^{(\text{Alt})}\otimes \left[\left(\C^d \right)^{\otimes N}\right]^{(\text{Alt})} \right)\\
        		&\oplus \tilde{\cH}^{(\text{Sym})},\\[1em]
    			\cH^{(\text{Alt})}&= \left( \left[\mathscr{H}^{\otimes N}\right]^{(\text{Sym})}\otimes \left[\left(\C^d \right)^{\otimes N}\right]^{(\text{Alt})} \right)\\
        		&\oplus \left( \left[\mathscr{H}^{\otimes N}\right]^{(\text{Alt})}\otimes \left[\left(\C^d \right)^{\otimes N}\right]^{(\text{Sym})} \right)\\
        		&\oplus \tilde{\cH}^{(\text{Alt})},
    		\end{split}
    	\end{equation}
see Eq. (\ref{app:hilbert_decomposition}) in App. \ref{app:subsec_hilbert}. Here, 
$\tilde{\cH}^{(\text{Sym})}, \tilde{\cH}^{(\text{Alt})}$ correspond to the product spaces of 
non-trivial symmetries which, when multiplied, give rise to totally symmetric and 
antisymmetric spaces respectively.  Specifically, the dimensions of $[(\C^d )^{\otimes 
N}]^{(\text{Sym})}$, and  $[(\C^d )^{\otimes N}]^{(\text{Alt})}$ are given by 
	\begin{equation}\label{eqn:dimensions_sym}
       	\begin{split}
        		\mathrm{dim}\left[\left(\C^d \right)^{\otimes N}\right]^{(\text{Sym})} &= \binom{d+N-1}{N},\\
        		\mathrm{dim}\left[\left(\C^d \right)^{\otimes N}\right]^{(\text{Alt})}  &= \binom{d}{N}.
        	\end{split}
    	\end{equation}
Note that if $d< N$, the antisymmetric subspace has dimension zero, meaning that it is not 
possible to construct a normalizable antisymmetric state. Just as in 
the two particle case, we can interpolate between bosonic and fermionic behavior of 
the energy degrees of freedom by controlling the auxiliary degrees of freedom associated with 
$\C^d$. In the following section, we will see how this control can be achieved via an 
energetic bias towards (or against) symmetric states of $(\C^d)^{\otimes N}$.

\section{Thermodynamics of Hamiltonian anyons}
\label{sec:thermodynamics}

In this section, we study the key thermodynamic properties of Hamiltonian anyons.  
We first derive the partition function and internal energy for an $N$-particle anyonic system subject to a Harmonic oscillator 
potential, demonstrating that such a system exhibits non-trivial first and second-order phase transitions. We later compare and 
contrast Hamiltonian anyons with the statistical anyons, as described in~\citeref{myers_21}.

\subsection{Partition function and internal energy}
\label{sec:Partition_function}
We consider a system of identical, non-interacting fermions in a one-dimensional harmonic trap 
with a symmetry-dependent energy bias as described by the following Hamiltonian
\begin{equation}\label{eqn:n_particle_hamiltonian}
        H = H_\mathrm{HO} \otimes \id_{\left(\C^d \right)^{\otimes N}} + \id_{\mathscr{H}^{\otimes N}}\otimes\nu\Pi_{\left(\C^d \right)^{\otimes N}}^\mathrm{(Alt)} +\mu\tilde{\Pi}.
\end{equation}
Here, $H_\mathrm{HO}$ denotes the Hamiltonian for a one-dimensional harmonic oscillator with 
$N$ particles
\begin{equation}
\begin{split}
    H_\mathrm{HO} & = \sum_{k=1}^N \left\{ \id_{\mathscr{H}}^{\otimes (k-1)} \otimes \left[\sum_{n=0}^\infty \hbar \omega \left( n + \frac{1}{2} \right) \ket{n}\bra{n}  \right] \right.\\
    & \left.  \otimes \ \id_{\mathscr{H}}^{\otimes (N-k)} \right\}.
    \end{split}
\end{equation}
$\Pi_{\left(\C^d \right)^{\otimes N}}^\mathrm{(Alt)}$ is the projector onto the 
antisymmetric subspace, $[(\C^d )^{\otimes N}]^{(\mathrm{Alt})}$; and $\tilde{\Pi}$ is the 
projector onto $\tilde{\cH}$, the product space of nontrivial symmetries. The parameter $\nu$ 
acts as an energy penalty for antisymmetric states of the auxiliary degree of freedom.
\changes{In the context of two spin-1/2 particles, a controlled bias of this type (known as singlet-triplet splitting) can be implemented in double quantum dots via the inter-dot distance or detuning \cite{he_05,stepanenko_12}; or else in organic optoelectronic materials by manipulating the spatial overlap of molecular orbitals \cite{chen_15}.} 
Similarly, the role of parameter $\mu$ is to penalize mixed-symmetry states. Here and 
throughout, we take $\mu$ to be much larger than any other energy scale in the system such 
that the population of $\tilde{\cH}$ can be entirely neglected. It will be assumed that when 
the system is weakly coupled to a large heat bath at inverse temperature $\beta$, it 
thermalizes to the Gibbs state $\tau = \frac{1}{Z}e^{-\beta H}$ \cite{binder_18}, with $Z = 
\tr[e^{-\beta H}]$ the partition function.  

\begin{figure}
    \includegraphics[width=\columnwidth]{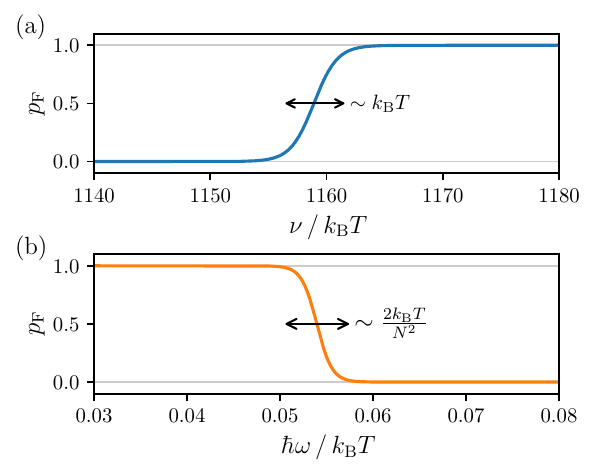}
    \caption{The fermionic subspace overlap, $\pf$, plotted as a function of $\nu/k_B T$ in panel (a) and $\hbar \omega/k_B T$ in panel (b). The system undergoes a transition from bosonic to fermionic statistics, as described by Eq. \eqref{eqn:pf}. In panel (a) we set  $\hbar\omega = k_\mathrm{B} T$, and the characteristic width of the sigmoid curve is $k_\mathrm{B} T$. In panel $(b)$ we set $\nu=0$; and the width of the transition is $\sim 2k_\mathrm{B} T / N^2$, becoming vanishingly narrow for large $N$. In both panels we have taken $N = d = 50$. Note the inflection point at $\pf =1/2$.}
\label{fig:anyonicity}
\end{figure}

As the overall state space belongs to $\cH^\mathrm{(Alt)}$, Eq. \eqref{eqn:hilbert_sym}, the 
projection of the Hamiltonian onto this space is
\begin{equation}\label{eqn:hamiltonian_decomp}
    \begin{split}
        H^\mathrm{(Alt)} &= H_\mathrm{HO}^\mathrm{(Alt)} \otimes \Pi_{\left(\C^d \right)^{\otimes N}}^\mathrm{(Sym)}\\
        &+ \left[H_\mathrm{HO}+\nu\id_{\mathscr{H}^{\otimes N}}\right]^\mathrm{(Sym)}\otimes\Pi_{\left(\C^d \right)^{\otimes N}}^\mathrm{(Alt)}\\
        &+ \mu\tilde{\Pi}^\mathrm{(Alt)}.
    \end{split}
\end{equation}
As the three terms of $H^\mathrm{(Alt)}$ in Eq. (\ref{eqn:hamiltonian_decomp}) act on mutually 
orthogonal subspaces, the Hamiltonian can straightforwardly be exponentiated as
\begin{equation}\label{eqn:ham_exponentiate}
    \begin{split}
        \left[e^{-\beta H}\right]^\mathrm{(Alt)} &= \left[e^{-\beta H_\mathrm{HO}}\right]^\mathrm{(Alt)} \otimes \Pi_{\left(\C^d \right)^{\otimes N}}^\mathrm{(Sym)}\\
        &+ e^{-\beta\nu}\left[e^{-\beta H_\mathrm{HO}}\right]^\mathrm{(Sym)}\otimes\Pi_{\left(\C^d \right)^{\otimes N}}^\mathrm{(Alt)},
        \end{split}
\end{equation}
where the mixed-symmetry term $e^{-\beta\mu}\tilde{\Pi}^\mathrm{(Alt)}$ has been neglected 
because $\beta\mu \gg 1$. Taking the trace of \eqnref{eqn:ham_exponentiate} we obtain the partition function  \begin{equation}\label{eqn:n_particle_partition}
     \begin{split}
         Z = \binom{d+N-1}{N} Z_\mathrm{F} + \binom{d}{N} e^{-\beta \nu} Z_\mathrm{B},
     \end{split}
 \end{equation}
where the binomial prefactors are the dimensions of $[(\C^d )^{\otimes N}]^{(\mathrm{Alt})}$ 
and $[(\C^d )^{\otimes N}]^{(\mathrm{Sym})}$ respectively. 
Equation \ref{eqn:n_particle_partition} is a linear combination of the partition functions for 
$N$ non-interacting fermions or bosons in a harmonic potential, 
$Z_\mathrm{F/B} = \tr\{[e^{-\beta H_\mathrm{HO}}]^\mathrm{(Alt/Sym)}\}$. The explicit 
expression of these are
 \begin{equation}\label{eqn:Z_n_fermion_boson}
 \begin{split}
    Z_\mathrm{F} &= e^{-\beta\hbar\omega\frac{N^2}{2}}\prod_{k=1}^N \frac{1}{1-e^{-\beta\hbar\omega k}},\\
    Z_\mathrm{B} &= e^{-\beta\hbar\omega\frac{N}{2}}\prod_{k=1}^N \frac{1}{1-e^{-\beta\hbar\omega k}},
    \end{split}
\end{equation}
where $\omega$ is the fundamental frequency of the oscillator \cite{schonhammer_00,mullin_02}. Tracing out the auxiliary degree 
of freedom one obtains the reduced thermal state, $\tau_{\mathscr{H}^{\otimes N}}$, of the harmonic oscillator degrees of 
freedom, which is a mixture of the fermionic and bosonic thermal states $\tau_\mathrm{F/B} = \frac{1}{Z_\mathrm{F/B}}[e^{-\beta 
H_\mathrm{HO}}]^\mathrm{(Alt/Sym)}$
\begin{equation}\label{eqn:tau_n_pos_pf}
     \begin{split}
         \tau_{\mathscr{H}^{\otimes N}} &= \binom{d+N-1}{N} \frac{\zf}{Z} \tau_{\mathrm{F}} + \binom{d}{N} e^{-\beta \nu} \frac{\zb}{Z} \tau_{\textrm{B}}\\
         &= p_{\mathrm{F}} \tau_\mathrm{F} + \left(1-p_{\mathrm{F}}\right) \tau_\mathrm{B},
     \end{split}
 \end{equation}
where the parameter $p_{\mathrm{F}}\in[0,1]$ reads
\begin{equation}\label{eqn:pf}
        p_{\mathrm{F}} = \frac{1}{1+ \exp[\frac{1}{2}N(N{-}1)\beta\hbar\omega - \beta\nu -h(d,N)]},
\end{equation}
and we have defined the shorthand $h(d,n) := \ln\binom{d+N-1}{N} - \ln\binom{d}{N}$ in the 
exponential. The parameter $\pf$ encapsulates the state's overlap with the antisymmetric 
subspace, interpolating from fully bosonic ($p_{\mathrm{F}} = 0$) to fermionic  ($p_{\mathrm{F}} = 1$) 
as $\nu$ is varied. Figure \ref{fig:anyonicity} shows the strictly monotonic behavior of $\pf$ with 
respect to parameters $\omega$ and $\nu$ 

The fermionic overlap $p_\mathrm{F}$ plays a crucial role also for the internal energy of the system. For an arbitrary state $\rho$, the internal energy is given by the expectation value of the Hamiltonian,  $U=\Tr[H^{\textrm{(Alt)}} \rho]$, where $H^{\textrm{(Alt)}}$ 
 as in Eq. (\ref{eqn:hamiltonian_decomp}). In the specific case of a thermal state,  $U = -\frac{\partial\ln{Z}}{\partial\beta}$ \cite{pathria2021}, and using Eq. (\ref{eqn:n_particle_partition}), we obtain
\begin{equation}\label{eqn:n_particle_internal_energy}
    \begin{split}
        U & = p_{\mathrm{F}} U_\mathrm{F} + \left(1-p_{\mathrm{F}}\right)\left(\nu + U_\mathrm{B}\right),
    \end{split}
\end{equation}
where the internal energies for bosons and fermions read 
\begin{equation}\label{eqn:fermion_bosons_internal_energy}
\begin{split}
    U_\mathrm{F} &= \hbar\omega\frac{N^2}{2} + \sum_{k=1}^N\frac{k\hbar\omega}{e^{k\beta\hbar\omega}-1},\\
    U_\mathrm{B} &= \hbar\omega\frac{N}{2} + \sum_{k=1}^N \frac{k\hbar\omega}{e^{k\beta\hbar\omega}-1},
\end{split}
\end{equation}
as shown in Ref. \cite{schonhammer_00}. Observe that $U_\mathrm{F}$ and $U_\mathrm{B}$ differ only in the ground state term. The difference corresponds to the fact that all $N$ bosons can occupy the harmonic oscillator ground state at energy $\hbar\omega/2$, whereas fermions must fill successive levels due to Pauli exclusion. For large $N$, and at low temperatures, the internal energies radically differ, a phenomenon sometimes referred to as \emph{Pauli energy} \cite{koch_23}. This is a signature of the relation between thermodynamics and exclusion statistics that drives thermodynamical behavior of systems of identical particles \cite{haldane_91,murthy_94}. 

\begin{figure*}
    \includegraphics[width=1.8\columnwidth]{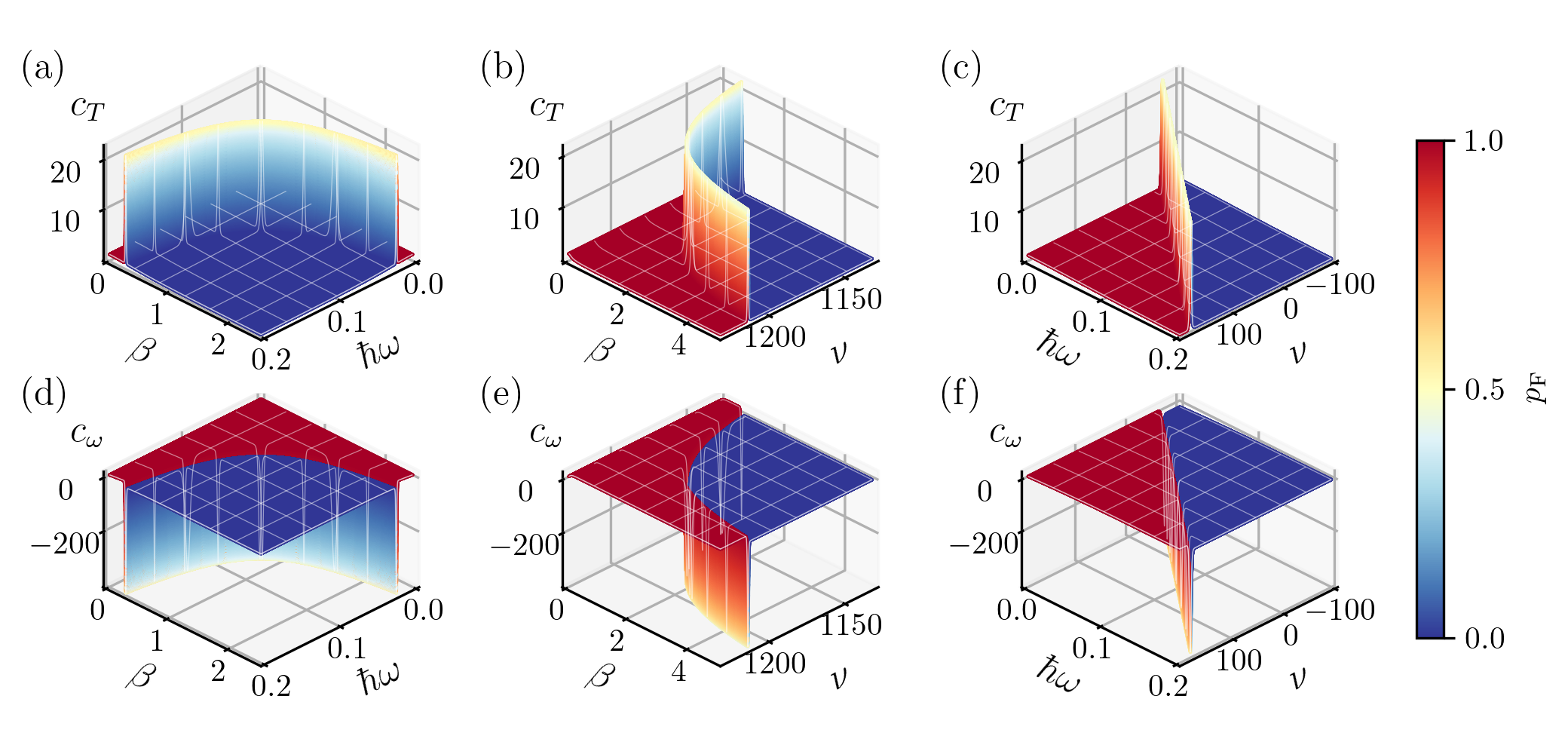}
    \caption{Diagrams of the heat capacity per particle, $c_T$ (panels (a)–(c)), and $c_\omega$ (panels (d)–(f)) for $N = d = 50$. The energies are in arbitrary units. Note that for $\omega$, there is an order of magnitude bigger jump in the derivative of $U$ than for $T$. The jump follows the expression given by Eq. (\ref{eqn:smallepsilon}) with $\varepsilon=0$, corresponding to equal overlap with the fermionic and bosonic subspaces.
}
\label{fig:3x3}
\end{figure*}

\subsection{Phase transitions}\label{subsec:phase}
We have seen that the internal energy of the $N$-particle ground state exhibits radically different 
scaling depending on exchange symmetry. One would therefore anticipate an 
abrupt change in the internal energy of Hamiltonian anyons at the transition between fermionic and 
bosonic statistics as characterized by the change in $\pf$, the overlap with the 
fermionic subspace. In this subsection, we demonstrate that the fermion-boson crossover constitutes a 
{\it bona fide} phase transition. A $k$-th order phase transition is characterized by a non-analyticity in the $k$-th derivative of the internal energy \cite{ehrenfest_33}.  

As seen in Fig. \ref{fig:anyonicity}, $\pf$ takes the form of a sigmoid curve centered at $\pf = 1/2$, 
where the parameters $\beta, \omega$ and $\nu$ satisfy 
    \begin{equation}
        \phi := \frac{1}{2} N(N-1) \beta\hbar\omega - \beta\nu- h(d,N)=0.
    \label{eqn:smallepsilon}
    \end{equation}
To check for phase transitions, we compute the first and second derivatives of internal energy, \eqnref{eqn:n_particle_internal_energy}, with respect to $T, \omega, \nu$, at the point where 
$\pf = 1/2$, see \appref{app:heat_capacity}. In the thermodynamic limit, $N\to\infty$, we find that the thermal capacities, $C_X=\partial U/\partial X, \, X\in\{T,\omega,\nu\}$, are given by
\begin{equation}\label{app:Capacities}
\begin{split}
        \lim_{N\to\infty}\frac{1}{N^2}C_T
         &= k_{\textrm{B}} \lim_{N\to\infty}
        \frac{h(d,N)^2}{4N^2},\\
        \lim_{N\to\infty}\frac{1}{N^2}C_\nu
        &= \lim_{N\to\infty}\frac{h(d,N)}{4N^2},\\
        \lim_{N\to\infty}\frac{1}{N^2}C_\omega
        &= \frac{\hbar}{4} - \frac{\hbar}
        {2}\lim_{N\to\infty} h(d,N).
\end{split}
\end{equation}
Note that the division by $N^2$ is necessary in order for the system to 
have a finite energy density in the thermodynamic limit. Observe that 
$C_\omega/ N^2$ diverges, whilst $C_T/N^2$ and $C_{\nu}/N^2$ remain bounded, see Eq. (\ref{app:limith}) in App. \ref{app:heat_capacities} for the limiting behavior of $h(d,N)$. 
Thus, the system undergoes a first-order phase transition as the trap 
frequency $\omega$ is varied.

The second derivatives of the internal energy can be shown to vanish at  
the inflection point $\pf=1/2$. However, by examining the behavior of the 
second derivatives $\varepsilon$ close on either side of that point, we 
find the following limiting behavior
\begin{equation*}
    \begin{split}
        \lim_{N\to\infty}\frac{1}{N^2}\frac{\partial^2 U}{\partial T^2} & = \lim_{N\to\infty}\frac{\varepsilon k_\mathrm{B}}{8T}\frac{\left[h(d,N)+\varepsilon\right]^3}{N^2}+\frac{1}{N^2}\UBbbeta,
\end{split}
\end{equation*}
\begin{equation}\label{app:second_derivatives}
\begin{split}
        \lim_{N\to\infty}\frac{1}{N^2}\Unnu& = \lim_{N\to\infty} \frac{\varepsilon\beta}{8}\left[\frac{h(d,N)+\varepsilon}{N^2}\right]-\frac{\beta}{2N^2}.
\end{split}
\end{equation}
As the second derivative with respect to temperature clearly diverges, it follows that the system experiences a second-order phase transition as the temperature is varied. Note, however, that no phase transition arises for the symmetry bias, $\nu$.

Figure \ref{fig:3x3} shows the specific heat per particle, $c_X = \frac{1}{N}\frac{\partial U}{\partial X}$, with 
$X\in\{T,\omega\}$ where we can clearly identify two regimes---determined by the relation between $T, \nu$ and $\omega$ at $\pf=1/2$---marked with an abrupt change in the specific heat. Note that this is a symmetry-driven phase transition similar to Bose-Einstein condensation or symmetry-breaking dynamical phase transitions \cite{manzano_14,manzano_18}.

\subsection{Comparison between Hamiltonian and statistical anyons}
\label{sec:comparison}
\begin{figure}[ht]
    \includegraphics[width=\columnwidth]{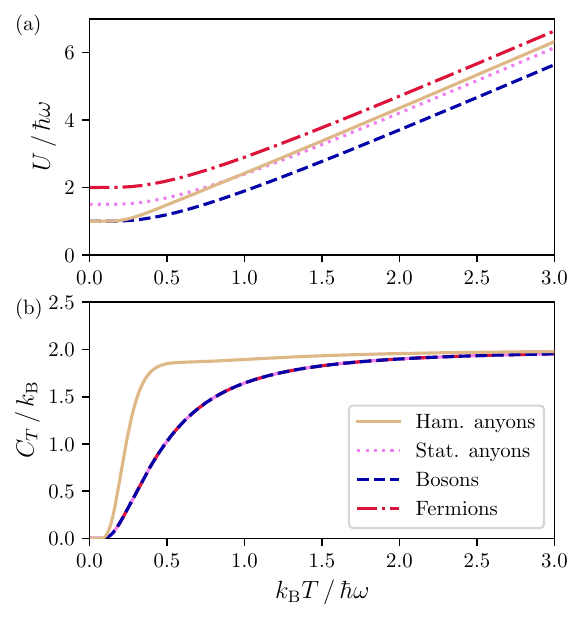}
    \caption{ Internal energy (a) and heat capacity (b) as a function of temperature, for a pair of fermions, bosons, statistical anyons with $k_\textrm{F}=0.5$, and Hamiltonian anyons with $\nu=0$. Statistical anyons are described by a fixed-weighting mean of fermions and bosons (which have identical heat capacities). Note that the pink, dotted line representing statistical anyons overlaps with the lines for bosons and fermions in panel (b). The internal energy of Hamiltonian anyons interpolates according to the temperature-dependent weighting $\pf$. This additional temperature dependence means that the heat capacity of Hamiltonian anyons can exceed that of fermions and bosons. These plots are for the case $N=2$, $d=2$.
    }
\label{fig:UC_comparison}
\end{figure}

In this subsection, we compare the thermal properties of Hamiltonian anyons with the statistical anyons proposed in 
\cite{myers_21}. Statistical anyons are based on an ensemble of noninteracting identical particles pairs whose symmetry can be 
manipulated, as in a Hong-Ou-Mandel effect \cite{hong_87}. For an ensemble composed of 
$N_\mathrm{F}$ fermionic systems and $N_\mathrm{B}$ bosonic ones, the properties of the system are characterized by the ratio 
$k_\mathrm{F} = \frac{N_\mathrm{F}}{N_\mathrm{F}+N_\mathrm{B}}$. In particular, the partition function $Z_\mathrm{SA}$ and 
internal energy $U_\mathrm{SA}$ for an ensemble of statistical anyon pairs are given by 
\begin{equation}\label{eqn:ZU_SA}
\begin{split}
    Z_\mathrm{SA} &= \left(Z_\mathrm{F}\right)^{k_\mathrm{F}} \left(Z_\mathrm{B}\right)^{(1-k_\mathrm{F})},\\
    U_\mathrm{SA} &= k_\mathrm{F} U_\mathrm{F} + (1-k_\mathrm{F}) U_\mathrm{B},
\end{split}
\end{equation}
where $Z_\mathrm{F/B}$ and $U_\mathrm{F/B}$ are defined in the previous section, see Eq. (\ref{eqn:fermion_bosons_internal_energy}). These equations contrast with their equivalent for Hamiltonian anyons, Eqs. (\ref{eqn:n_particle_partition}) and (\ref{eqn:n_particle_internal_energy}).

Both Hamiltonian and statistical anyons have similar thermodynamic properties, with $p_\mathrm{F}$ and $k_\mathrm{F}$ playing a similar role. The main difference between the two models is the dynamic character of Hamiltonian anyons, as the parameter $\pf$ depends on the variables $\omega, \nu$ and $\beta$, in contrast to statistical anyons where $k_\mathrm{F}$ is fixed.
We shall see in Section \ref{sec:engines} that the adaptive nature of Hamiltonian anyons confers novel advantages for heat engines. 

In Fig. \ref{fig:UC_comparison}, we compare the internal energy, Fig. \ref{fig:UC_comparison}(a), and heat capacity $(C_T)$, Fig. \ref{fig:UC_comparison}(b), for statistical and Hamiltonian anyons as well as for bosons and fermions, as a function of temperature. The comparison is made for $k_F=0.5$ and $\nu=0$.
The general trend of the system energy is the same for all cases, increasing when $T$ increases or when $\omega$ decreases.  The behavior of both Hamiltonian and statistical anyons interpolates between that of purely bosonic and fermionic ensembles. However, due to their dynamical nature, and the presence of non-trivial phase transitions, Hamiltonian anyons exhibit a much higher heat capacity that of all other mediums.  This phenomenon can be exploited for the construction of engines with increased efficiency and power as we now demonstrate.

\section{Quantum engines based on symmetry control}
\label{sec:engines}

Particle exchange symmetry is known to affect the performance of heat engines at low temperatures 
\cite{myers_20,myers_21}, and 
work extraction cycles based on manipulating particle statistics have recently been demonstrated in ultra-cold 
gases~\cite{koch_23}. In this section we demonstrate how the ability to dynamically vary the statistical properties of 
Hamiltonian anyons can be exploited for novel advantages in quantum heat engines. Two types of cycle will be considered: a 
Stirling cycle in which the conventional compression and expansion strokes are replaced by 
fermionization and bosonization of the working medium (\secref{sec:Stirling}), and a finite-time Otto cycle where the medium 
passively transitions between fermionic and bosonic statistics during heating and cooling (\secref{sec:Otto}). 

Before analyzing the performance of these cycles, let us briefly review some basic concepts from quantum thermodynamics and 
define the notions of heat and work that we will use.  We assume that parameters $\omega$ and $\nu$ can be externally 
controlled, and that the system may be connected at different times with different heat baths, each characterized by an inverse
temperature $\beta$. \changes{In this regard, electrostatically-defined quantum dots provide a promising platform for implementation \cite{campbell_26}: In a double dot system, both the natural frequency of the confining potential and the energy splitting $\nu$ between singlet and triplet states can be varied via gate electrodes \cite{he_05,stepanenko_12}; and interaction between the system and reservoir electrodes can be effectively turned on and off via tuneable tunneling barriers \cite{wenz_19}.}

The cycles considered here each consist of two \emph{driving} strokes and two \emph{thermalization} strokes. During a driving stroke, either $\omega$ or $\nu$ is varied, causing the system to evolve according to a time-dependent Hamiltonian while in contact with a fixed-temperature bath. Conversely, in a thermalization stroke, the system is disconnected from one heat bath and connected to another, whereupon it is allowed to equilibrate to the Gibbs state at the new temperature, with the Hamiltonian remaining fixed throughout.

The rates of work and heat transfer will be defined according to~\cite{alicki_79}
\begin{equation}
    \begin{split}
        \dot{W} &= -\tr\left(\rho\dot{H}\right),\\
        \dot{Q} &= \tr\left(\dot{\rho}H\right),
    \end{split}
\label{eq:W_and_Q}
\end{equation}
using the convention that positive work is done by the system against the driving field, and positive heat is absorbed into the 
system from the bath. We note that these definitions are consistent with the first law of thermodynamics, 
$\dot{U} = \dot{Q} - \dot{W}$ \cite{kosloff13,binder_18,strasberg2022quantum}.

During the thermalization strokes, the Hamiltonian is unchanging, i.e., $\dot{H} = 0$, so changes to the internal energy of the  
system are purely heat. During driving strokes, the work and heat depend on how the stroke is performed. In order to simplify 
the analysis we shall consider two idealized driving schemes: adiabatic fast driving and quasistatic isothermal driving.  In 
adiabatic fast driving, the Hamiltonian is switched from $H_i$ to $H_f$ much faster than the system's evolution, so 
that the state remains unchanged, i.e., $\dot{\rho}=0$. Consequently, $Q=0$ and the change in internal energy of the system is 
purely work, $W = \tr(\rho H_i) - \tr(\rho H_f)$. On the other hand, in quasistatic isothermal driving, the 
Hamiltonian is changed from $H_i$ to $H_f$ at a rate which is slow in comparison to thermalization, resulting in 
the system remaining in the instantaneous Gibbs state $\tau[\beta,H(t)]$ at all times. In this case work is equal to the change 
in free energy, i.e. 
\begin{equation}\label{qswork}
    W = \frac{1}{\beta}\ln \left[\frac{Z(\beta,H_f)}{Z(\beta,H_i)}\right],
\end{equation}
where the partition function $Z$ is as in Eq. \eqref{eqn:n_particle_partition}. This is accompanied by a nontrivial heat 
transfer, which can be determined using the first law as $Q = U(\beta,H_f) - U(\beta,H_i) + W$.

\begin{figure}[t]
    \includegraphics[width=\columnwidth]{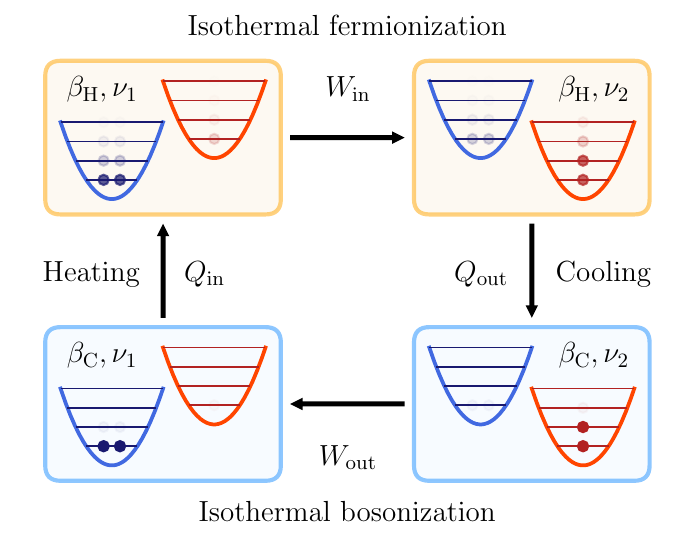}
    \caption{Schematic diagram of a modified Stirling cycle based on \emph{fermionizing} and \emph{bosonizing} the particle 
    statistics. A pair of Hamiltonian anyons has access to two ladders of energy eigenstates, one with bosonic exchange symmetry 
    (blue) and one fermionic (red), which are offset by a variable energy bias $\nu$. As both particles can occupy the bosonic 
    ground state, the internal energy is lower than for fermions which are subject to Pauli exclusion. By lowering $\nu$, so 
    that the particles have access to bosonic states, energy can be extracted as work. Moreover, by bosonizing at one 
    temperature and fermionizing at another, a net output of work can be achieved in a cycle.}
    \label{fig:stirling_sketch}
\end{figure}
\subsection{\label{sec:Stirling} Stirling Cycle}
In a classical heat engine, a piston varies the volume of a cylinder of gas in order to extract useful work. A quantum 
equivalent would vary the fundamental frequency $\omega$ of the confining potential
\cite{zheng_16,son_21,tejero_24,tejero_preprint}. Instead, we here consider actively driving the statistical properties of a 
medium of Hamiltonian anyons via the parameter $\nu$, as shown in \figref{fig:stirling_sketch}, while holding $\omega$ fixed 
throughout.   
Starting in a thermal state at temperature $\beta_\mathrm{H}$ and $\nu=\nu_1$, the parameter $\nu$ is quasistatically and 
isothermally driven to $\nu_2$ while the system remains in equilibrium with the hot bath. The system is then connected to the 
cold bath and thermalizes to $\beta_\mathrm{C}$ with $\nu$ fixed. The parameter $\nu$ is then isothermally driven back to 
$\nu=\nu_1$, and finally the cycle is completed by allowing the system to thermalize with the hot bath at $\beta_\mathrm{H}$.

In general, when the bias is isothermally driven from $\nu_i$ to $\nu_f$, the work extraction is given by
\begin{equation}\label{nuwork}
    W = \frac{1}{\beta}\ln \left[\frac{p_\mathrm{F}(\beta,\nu_i)}{p_\mathrm{F}(\beta,\nu_f)}\right], 
\end{equation}
where we have used the fact that $\pf = \binom{d+N-1}{N} \frac{Z_\mathrm{F}}{Z}$ with 
$Z_\mathrm{F}$ independent of $\nu$. Eq. \eqref{nuwork} makes evident the strikingly direct relationship between thermodynamic work and the change in particle exchange symmetry, as characterized by $\pf$. As $p_\mathrm{F}$ is a monotonically increasing function of $\nu$ (see Fig. \ref{fig:anyonicity}), lowering $\nu$ (bosonizing) necessarily results in positive work extraction, while raising $\nu$ (fermionizing) constitutes work done on the system. See Fig. \ref{fig:stirling_sketch} for a physical interpretation. 

\begin{figure}[t]
    \includegraphics[width=0.85\columnwidth]{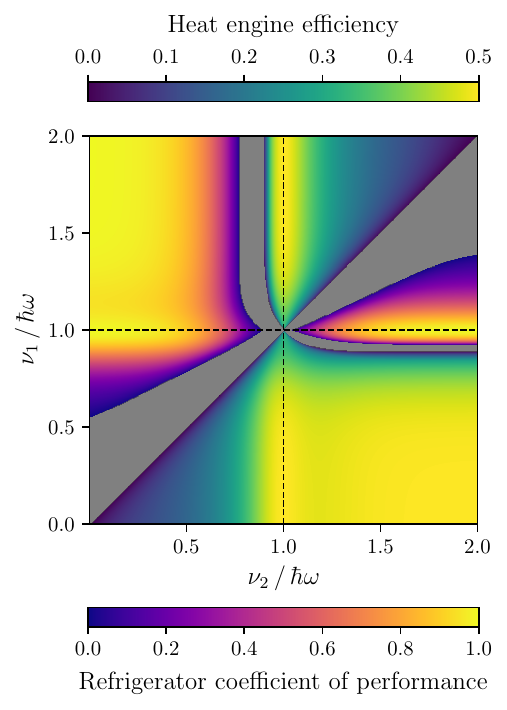}
    \caption{
    Performance of a modified Stirling cycle with a pair of Hamiltonian anyons as a working substance $(N=d=2)$. All energies are in units of $\hbar\omega$, which 
    remains fixed throughout the cycle, and the bath inverse-temperatures are $\beta_\mathrm{H} = 10\hbar\omega$ and $\beta_\mathrm{C} = 20\hbar\omega$ such 
    that $\eta_\mathrm{Carnot} = 0.5$. Thermodynamic efficiency is plotted in regimes where the cycle extracts positive work; and refrigerator coefficient of performance is plotted where the cycle removes heat from the cold bath. In the upper-right and bottom-left quadrants, the engine extracts work whenever $\nu_\mathrm{1} > \nu_\mathrm{2}$ and $\nu_\mathrm{1}<\nu_\mathrm{2}$ respectively. The cycle approaches reversible efficiency in some regimes, notably in the lower-right, where the system is driven far either side of the fermion-boson transition. In the regimes which are shaded gray, the cycle functions neither as a heat engine nor a refrigerator.}
    \label{fig:stirling_performance} 
\end{figure}

The net work extraction during the cycle is given by summing both isothermal strokes
\begin{equation}\label{cyc_work}
    W_\mathrm{cyc} = \frac{1}{\beta_\mathrm{H}}\ln\left[\frac{p_\mathrm{F}(\beta_\mathrm{H},\nu_\mathrm{1})}{p_\mathrm{F}(\beta_\mathrm{H},
    \nu_\mathrm{2})}\right] +  \frac{1}{\beta_\mathrm{C}}\ln\left[\frac{p_\mathrm{F}(\beta_\mathrm{C},\nu_\mathrm{2})}{p_\mathrm{F}(\beta_\mathrm{C},
    \nu_\mathrm{1})}\right].
\end{equation}
Due to the non-trivial temperature dependence of $p_\mathrm{F}$, $W_\mathrm{cyc}$ is not guaranteed to be positive, and the 
different regimes of the cycle's usefulness are represented in Fig. \ref{fig:stirling_performance}. Perhaps surprisingly, 
it can be advantageous to perform the work extraction stroke at the colder isotherm. \changes{From an intuitive perspective, this is because when driving $\nu$ (modifying the energy of fermionic states) the work increment is $\delta W = - \pf(\nu,\beta)\, \delta \nu$; and so clearly it is advantageous to perform the work extraction stroke when $\pf \approx 1$. Since the fermion-boson transition becomes sharper at colder temperatures -- see Fig.\eqref{fig:anyonicity} -- this leads to a more extremal value of $\pf$ and hence greater work extraction, sometimes enough to outweigh the penalty of raising $\nu$ at a warmer temperature. 
}
See Appendix \ref{app:engines} 4 for a more formal treatment.

A key metric of performance is the cycle's thermodynamic efficiency, given by $\eta = W_\mathrm{cyc} /Q_\mathrm{H}$, where 
$Q_\mathrm{H}$ is the net heat absorbed from the hot bath. \changes{Since the model does not include a regenerator system to recover waste heat, this Stirling-type cycle does not attain reversible efficiency except in the limits of cold temperature and complete bosonization/fermionization, as discussed below.} By considering the total change of internal energy and work done 
while in contact with the hot bath, we can write an expression for $Q_\mathrm{H}$ using the first law
\begin{equation}\label{Qh}
    Q_\mathrm{H} = U(\beta_\mathrm{H},\nu_2) - U(\beta_\mathrm{C},\nu_1) + \frac{1}{\beta_\mathrm{H}}\ln \left[\frac{p_\mathrm{F}(\beta_\mathrm{H},\nu_1)}
    {p_\mathrm{F}(\beta_\mathrm{H},\nu_2)}\right],
\end{equation}
where $U$ is given by Eq. \eqref{eqn:n_particle_internal_energy}. Figure \ref{fig:stirling_performance} shows a plot of 
efficiency as a function of the parameters $\nu_1, \nu_2$.  

In the limiting cases $p_\mathrm{F}(\beta,\nu_2){\to} 1$ and $p_\mathrm{F}(\beta,\nu_1){\to} 0$, where $\nu$ is driven though a 
sufficiently large range of values for \emph{complete} fermionization and bosonization, the net work and heat of the cycle are 
(see \appref{app:engines})
\begin{equation}\label{limitingwork}
    \begin{split}
        W_\mathrm{cyc} &\to \left(\frac{1}{\beta_\mathrm{H}} -  \frac{1}{\beta_\mathrm{C}}\right) h(d,N),\\
        Q_\mathrm{H} &\to \frac{1}{\beta_\mathrm{H}} h(d,N) + \sum_{k=1}^N \left( \frac{k\hbar\omega}{e^{k\beta_\mathrm{H}\hbar\omega}-1} - \frac{k\hbar\omega}{e^{k\beta_\mathrm{C}\hbar\omega}-1} \right).
    \end{split}
\end{equation}
Under the further assumption that $\beta\hbar\omega\gg 1$ for both temperatures (i.e. in very cold regimes where particle statistics matter the most), the second term in $Q_\mathrm{H}$ vanishes, and the cycle attains the ideal Carnot efficiency
\begin{equation}
    \eta\to 1-\frac{\beta_\mathrm{H}}{\beta_\mathrm{C}}.
\end{equation}
Additionally, there exist regions of $\nu_1$ and $\nu_2$ where the cycle acts as a 
refrigerator, absorbing a net positive amount of heat $Q_\mathrm{C}$ from the cold bath at the cost of consuming work ($W_\mathrm{cyc}\leq 0$). For such cases, the coefficient of 
performance $ Q_\mathrm{C} / |W_\mathrm{cyc}|$ is plotted as a function of $\nu_1$ and $\nu_2$ in 
\figref{fig:stirling_performance}.

\subsection{Finite-time Otto cycle}\label{sec:Otto}
\begin{figure}
    \includegraphics[width=\columnwidth]{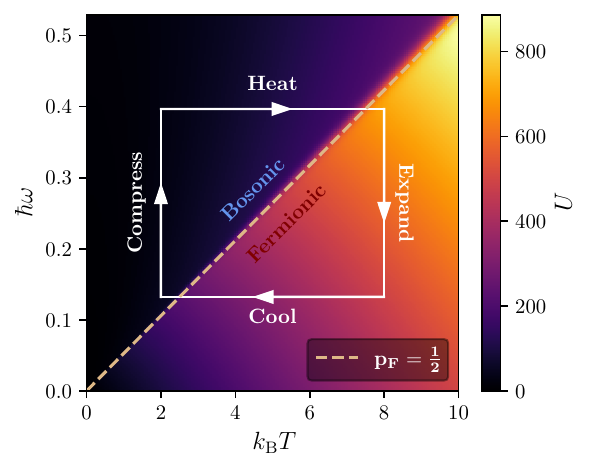}
    \caption{Sketch of the Otto cycle, overlaid on a heatmap of internal energy $U$ against temperature $T$ and trap frequency $\omega$ for  $N=d=50$ and $
    \nu=0$ (all energies in arbitrary units). By crossing the anyonic transition during the heating and cooling strokes, the work output during expansion can be 
    boosted by the effective repulsion between particles in the fermionic phase; while compression is performed at reduced cost in the bosonic phase.}
    \label{fig:otto_sketch}
\end{figure}

Here, we consider an Otto cycle which involves a temperature-induced transition in a working substance of 
Hamiltonian anyons. Unlike the Stirling engine, the strokes in an Otto cycle occur in finite time, allowing us to study 
enhancements to an engine's \emph{power}. The cycle acts between two heat baths at temperatures 
$\beta_\mathrm{H}<\beta_\mathrm{C}$ with the strokes implemented by varying the frequency of the trap as shown in 
\figref{fig:otto_sketch}. \changes{Here we use a sudden quench model as detailed in \cite{watson_25_2,watson_25}, which permits considerable simplification of the analysis and leads to results which are independent of any particular model of thermalization dynamics.}

Starting in the thermal state $\tau(\beta_\mathrm{H},\omega_{1})$, the trap frequency is quickly and adiabatically switched from 
$\omega_{1}$ to a smaller value $\omega_{2}$---corresponding to an expansion of the confining potential---during which the state 
undergoes no evolution. The system is then placed in contact with a cold bath and thermalizes to 
$\tau(\beta_\mathrm{C},\omega_2)$. Following this, the frequency is instantaneously switched back from $\omega_2$ to $\omega_1$, 
compressing the trap, with the system remaining in the state $\tau(\beta_\mathrm{C},\omega_\mathrm{2})$. The cycle is completed 
by placing the system in contact with the hot bath, allowing it to thermalize back to its original state 
$\tau(\beta_\mathrm{H},\omega_1)$. 

The value of $\nu$ remains fixed throughout, which we take to be $\nu = 0$ for simplicity. Consequently, as can be seen from Eq. 
\eqref{eqn:n_particle_hamiltonian}, the eigenvalues of the overall Hamiltonian coincide with those of the harmonic oscillator and are
therefore proportional to $\omega$. By extension, the internal energy following the adiabatic expansion is simply changed by a 
factor $\omega_2 / \omega_1$, since the state remains fixed during driving. The work output of the expansion stroke is equal to 
the resulting reduction in internal energy given by
    $(1-\frac{\omega_2}{\omega_1}) U(\beta_\mathrm{H},\omega_1)$, 
where $U$ is as in Eq. \eqref{eqn:n_particle_internal_energy}. Similarly, the internal energy changes by a factor $\omega_1 / \omega_2$ during the compression stroke so that the work done \emph{on} the system is 
$(\frac{\omega_1}{\omega_2} - 1) U(\beta_\mathrm{C},\omega_2)$.
Therefore the net work extraction per cycle is
\begin{equation}\label{otto_work}
    \begin{split}
        W_\mathrm{cyc} = \left(1-\frac{\omega_2}{\omega_1}\right) U(\beta_\mathrm{H},\omega_1) - \left(\frac{\omega_1}{\omega_2} - 1\right) U(\beta_\mathrm{C},\omega_2).
    \end{split}
\end{equation}
On the other hand, the heat absorbed from the hot reservoir is equal to the increase in internal energy during the heating stroke
\begin{equation}\label{otto_heat}
    Q_\mathrm{H} = U(\beta_\mathrm{H},\omega_1) - \frac{\omega_1}{\omega_2} U(\beta_\mathrm{C},\omega_1),
\end{equation}
The same quantities can be evaluated for a working substance of pure 
bosons or fermions by replacing $U$ in the above with $U_\mathrm{F}$ or $U_\mathrm{B}$ as given in Eq. \eqref{eqn:fermion_bosons_internal_energy}.
\changes{The efficiency of the cycle is given by the quotient of Eqs.(\ref{otto_work},\ref{otto_heat})
and it is shown in App.\eqref{app:engine2} that this expression simplifies to
\begin{equation}
    \eta = 1 - \frac{\omega_2}{\omega_1},
\end{equation}
regardless of whether the working substance consists of bosons, fermions or anyons; and independent of the reservoir temperatures or particle number. The reason this cannot exceed the Carnot efficiency is that $W_\mathrm{cyc} < 0$ wherever $\omega_2/\omega_1 < \beta_\mathrm{H}/\beta_{C}$ (i.e. the cycle ceases to function as an engine). Nonetheless, Carnot efficiency can be approached as this inequality is saturated, at the expense of vanishing work (see App.\ref{app:engine2}).}

While anyons offer no gain in efficiency in this model, the work output $W_\mathrm{cyc}$ can be significantly boosted when operating across the anyonic transition. \changes{Similar power enhancements have been predicted for cycles operating across a Bose-Einstein condensation transition \cite{Myers_22,Eglinton_23}.}

To see the origin of this, note that the ground-state contribution to 
the internal energy is $\hbar\omega N^2 / 2$ for fermions, as opposed to $\hbar\omega N / 2$ for bosons.  Hence, for large $N$, the same $\Delta \omega$ can entail a much larger work transfer for fermions than bosons. We can exploit this fact by tuning the cycle such that expansion 
takes place in the fermionic regime whilst compression occurs in a bosonic regime. For given reservoir temperatures $\beta_\mathrm{C}$ and $\beta_\mathrm{H}$, 
this situation can be engineered by choosing $\omega_1$ and $\omega_2$ such that $p_\mathrm{F}(\beta_\mathrm{C},\omega_2)\approx 0$ and $p_\mathrm{F}(\beta_\mathrm{H},\omega_1)\approx 1$ (see \figref{fig:otto_sketch}).

\begin{figure}
    \includegraphics[width=\columnwidth]{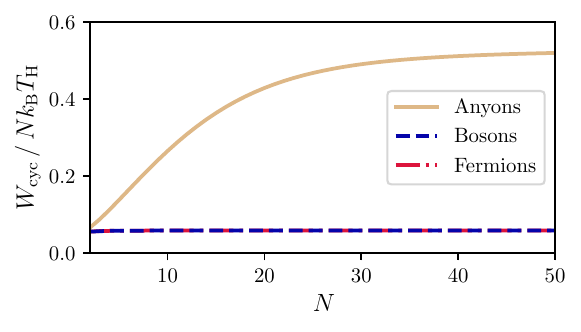}
    \caption{Performance comparison for a fast-switching Otto cycle with fermions, bosons, and Hamiltonian anyons as a working substance. Work per particle per cycle $W_\mathrm{cyc}/Nk_\textrm{B}T_\textrm{H}$ \changes{(effectively the power-density)} is plotted against the number of particles $N$ in the working substance. \changes{While fermions and bosons exhibit identical performance}, Hamiltonian anyons display a significant boost to work extraction owing to the dynamical transition in particle statistics, with the advantage becoming more pronounced as the number of particles grows. Here, the ratio between reservoir temperatures is taken to be $\beta_\mathrm{C}/ \beta_\mathrm{H}=2$, and the two trap frequencies are chosen such that $\phi(\beta_\mathrm{H},\omega_1)=-0.1$ and $\phi(\beta_\mathrm{C},\omega_2)=+0.1$ \changes{where $\phi$ is as in Eq. \eqref{eqn:smallepsilon}. Fixing the compression ratio this way determines the efficiency as $\eta = 1 - \frac{1.1\times\beta_\mathrm{H}}{0.9\times\beta_\mathrm{C}} = \frac{7}{18}$, regardless of the system size or composition.} We set $d=N$ and $\nu =0$.}
    \label{fig:otto_performance}
\end{figure}

Figure \ref{fig:otto_performance} compares power density,  
$W_\mathrm{cyc} / N$, between bosonic, fermionic and Hamiltonian anyon 
mediums for a fast-switching Otto engine. Unlike the statistical anyons in \cite{myers_21}, 
whose thermal properties interpolate between those of bosons 
and fermions, Hamiltonian anyons surpass both bosonic and fermionic working substances. The 
advantage can be directly attributed to the transition in particle statistics throughout the 
cycle, the effect of which becomes more pronounced as the number of particles of the working 
medium increases \footnote{While $W_\mathrm{cyc}$ does increase with $N$ for fermions and bosons, it tends to a limit extremely quickly, leading to the impression of a fixed value in Fig.\eqref{fig:otto_performance}.}. 

\section{Simulation on a quantum computer}
\label{sec:simulation}
\begin{figure}[t]
    \includegraphics[width= \columnwidth]{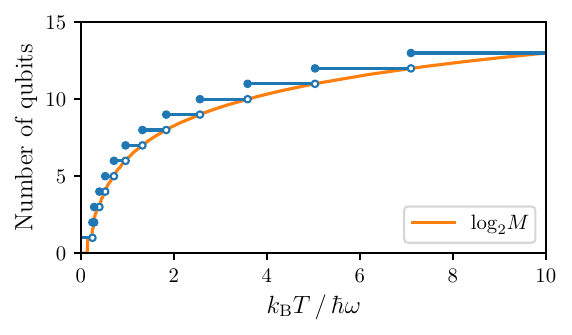}
    \caption{Number of qubits required to represent the energy levels which account for 99.9\% population in the thermal state, plotted against temperature for $N=d=2$. The parameter $\nu$ is adjusted so that $\pf=1/2$ at all $T$ values. The blue lines mark the ratio of temperature to energy that can be captured by a given integer number of qubits.}
\label{fig:qubit_requirement}
\end{figure}

In this section, we briefly discuss how to simulate Hamiltonian anyons on digital quantum 
computers. The main idea is to divide the qubits of a quantum processor into two 
sets. A first set comprises those that are used for simulating the energy degrees of freedom, and a second one with those that are used for simulating the spin degrees of freedom. 
For particles with spin $j$, so that $d=2j+1$, the latter corresponds to $\log_2 d$ 
qubits.  For the energy degrees of freedom, the number of 
qubits used depends strongly on how many eigenstates of the Harmonic oscillator we wish to simulate. Figure \ref{fig:qubit_requirement} plots the number of qubits 
needed to encode the energy levels that account for 
99.9\% of the population present in the thermal state of two particles as a function 
of temperature. Setting an upper bound $M$ on the total energy eigenstates implies 
$\log_2 M$ qubits.  Hence, to simulate $N$ such particles requires a total of $\log_2 M+ N \log_2 d$ qubits. 

\begin{figure*}[ht]
\centering
        \includegraphics[width=0.8\textwidth]{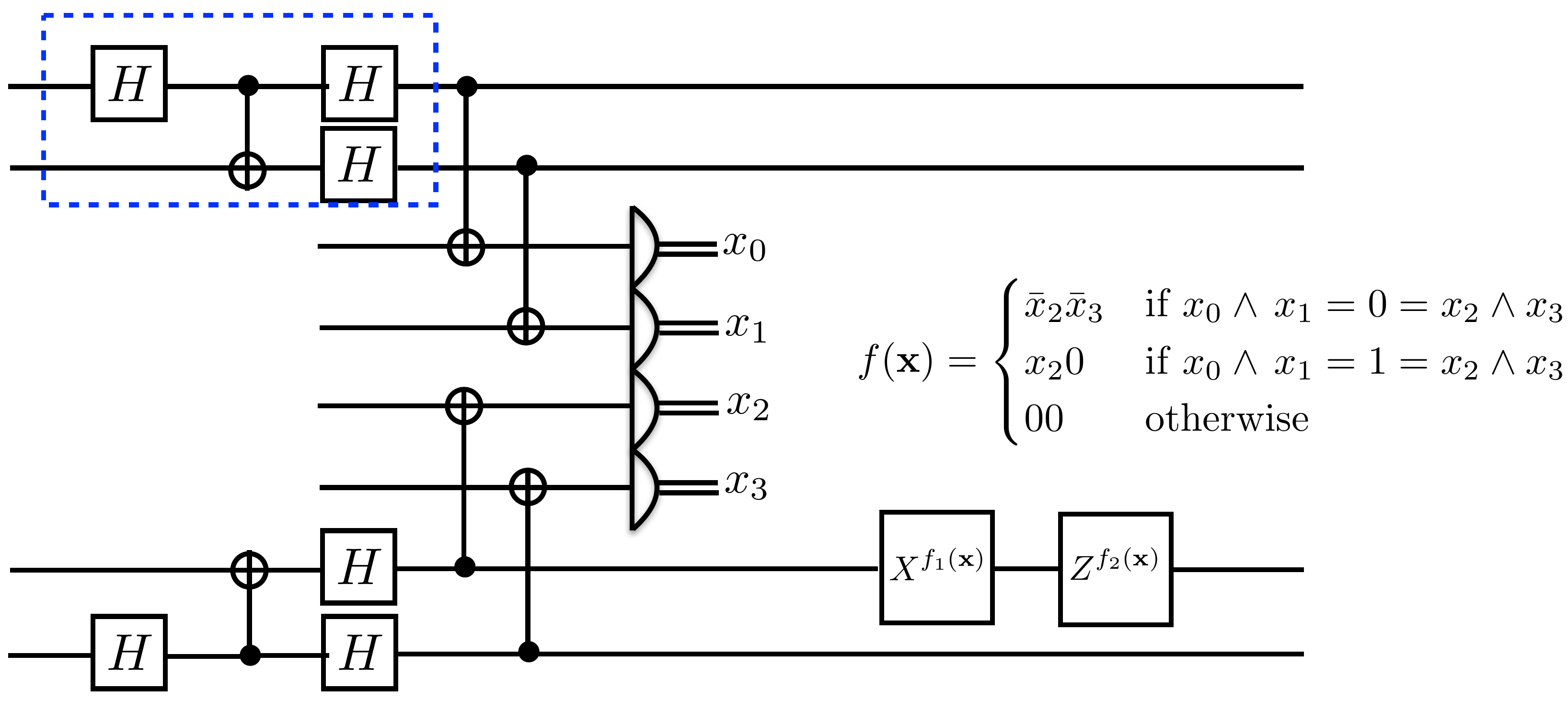}
        \caption{A quantum circuit implementation for a pair of Hamiltonian anyons with $M=2, d=2$ prepared such that their global states belongs to $\cH^{(\mathrm{Alt})}$. 
        The top two qubits encode the energy spectrum of the particles whilst the bottom two the spin degrees of freedom. The subcircuit in blue performs a Bell basis measurement whose outcome is then imprinted onto the state of four auxiliary qubits (two for each Bell basis measurement). The outcome $x_0=x_1=1$ (and likewise for $x_2$ and $x_3$) corresponds to the totally antisymmetric state.  Depending on the measurement outcomes, $x_k\in\{0,1\}$, the function $f: \{0,1\}^4\to\{0,1\}^2$, determines  how the spin degrees of freedom are modified so that if the energy eigenstates are symmetric (antisymmetric) the corresponding spin states are antisymmetric (symmetric). The input states to the algorithm can be arbitrary, and the output states are then fed into an algorithm that prepares the thermal state for the Hamiltonian of \eqnref{eqn:n_particle_hamiltonian}.}
\label{fig:circuit}
\end{figure*}

Qubits in a quantum computer do not inherently possess the symmetry properties required for simulating Hamiltonian anyons. Hence, these have to be encoded beforehand.
In \figref{fig:circuit} we show how to prepare on a quantum computer a totally antisymmetric state for the case where $M=d=2$, i.e., a pair of spin-$1/2$ particles, whose energy spectrum is limited to three distinct energy eigenvalues.  The circuit consists of preparing a pair of the Bell basis states 
    \begin{equation}
        \ket{\Phi_{ij}}=\left(X^i Z^j\otimes \one\right)\frac{1}{\sqrt{2}}\sum_{k=0}^1\ket{k}\otimes\ket{k} ,
    \label{eq:Bell_states}
    \end{equation}
where $X$ and $Z$ are the Pauli operators.  Note that for $i=j=1$ the state $\ket{\Phi_{11}}$ is totally antisymmetric. For each pair of Bell states the classical labels $ij$ are imprinted onto the state of two auxiliary qubits 
which are then measured to produce the classical bit string ${\bf x}\in\{0,1\}^4$.  Observe that the logical AND operator between the bits $x_0$ and $x_1$ (and likewise between bits $x_2$ and $x_3$) uniquely identifies the totally symmetric and antisymmetric states, i.e. $x_0\wedge x_1=1$ if and only if the state is antisymmetric.  If $x_0\wedge 
x_1\neq x_2\wedge x_3$ then the global state of the four qubits belongs to the totally antisymmetric subspace.  If, however, $x_0\wedge x_1 = x_2\wedge x_3$ then the states are 
either both symmetric or antisymmetric.  In the former case (corresponding to $x_0\wedge x_1 = 0$) we apply $X^{\bar{x}_2} Z^{\bar{x}_3}$, where $\bar{x}$ denotes the negation 
of $x$, to the third qubit which ensures that the last two qubits in \figref{fig:circuit} are in the state $\ket{\Phi_{11}}$.  If $x_0\wedge x_1 = 1$ we simply apply $X^{x_2}$ to the 
third qubit ensuring that the last two qubits in \figref{fig:circuit} are in the symmetric state $\ket{\Phi_{10}}$.  

Next, we need to prepare the thermal state of \eqnref{eqn:tau_n_pos_pf}. This is accomplished by implementing a quantum analogue of the classical Markov Chain Monte Carlo algorithm \cite{Temme2011}.  This implementation proceeds via simulation of the following dynamical operator over a generic state $\rho$~\cite{chen_quantum_2023}
    \begin{equation}
        \begin{split}
    \cL_\beta[\rho]&:=-\dfrac{i}{\hbar}\left[H,\rho\right]+\sum_{k\in K}\int_{-\infty}^{\infty} \gamma(x)\left[\hat{L}_k(x)\rho\hat{L}^\dagger_k(x)\right.\\
        &\left.-\frac{1}{2}\left\{ \hat{L}^\dagger_k(x)\hat{L}_k(x),\rho\right\}\right]d x,
        \end{split}
    \label{eq:Lindblad_thermal_state}
    \end{equation}
where $H$ is the $N$-particle Hamiltonian, see \eqnref{eqn:n_particle_hamiltonian}. The coefficients $\gamma(x)$ are the transition weights, which must satisfy the detailed-balance condition $\frac{\gamma(x)}{\gamma(-x)}=e^{-\beta x}$, and $\hat{L}_k(x)$ is the weighted Fourier transform of the $k$-th jump operator, $L_k(t)$, i.e., 
\begin{equation}
\hat{L}_k(x)=\frac{1}{\sqrt{2\pi}}\int_{-\infty}^{\infty} e^{-i x t} f(t) L_k(t)d t.
\label{eq:FT_jumps}
\end{equation}
Here $f(t)$ is a real, square-integrable filter function and the set of jump operators $\{L_k(t)\}$ is closed under adjoints and satisfies $\lVert\sum_{k\in K} L^\dagger_k(t) L_k(t) \rVert\leq 1$. For a suitably chosen filter function, \citeref{chen_quantum_2023} shows that the distance between the fixpoint state of $\cL_\beta$ and the true thermal state is less than $\cO\left(\sqrt{\frac{\beta}{t}}t_{\mathrm{mix}}(\cL_\beta)\right)$, where $t_{\mathrm{mix}}(\cL_\beta)$ is the mixing time of the dynamics~\footnote{The mixing time of any open system dynamics is defined as the time it takes for any pair of initial states to become indistinguishable.}. 

The implementation on a quantum processor proceeds by first discretizing the Fourier transform  and performing a block encoding of the dynamics into a unitary operator with the help of auxiliary qubit registers (see Lemma I.1 in \cite{chen_quantum_2023}). The resource scaling of the implementation is linear in time, and requires a bounded number of auxiliary qubit registers that depends on the block encoding, the total number $K$ of jump operators and the discretization of the Fourier transform the latter being the most resource consuming step.

\section{Conclusions}
\label{sec:conclusions}
In this paper, we have introduced a novel model of statistical anyons. We have shown that the behavior of our model  
interpolates non-trivially between that of fermions and bosons. Our model exhibits rich thermodynamic properties, including phase transitions, that are 
controllable by external parameters. This dynamic behavior stands in sharp contrast to the more static properties of topological and statistical anyons, which lack such phase transitions. We have shown how Hamiltonian anyons can be controlled in order to construct engine cycles which exhibit enhancements in both their efficiency and power output as compared to a working medium of  bosons, fermions, or statistical anyons. We have also demonstrated that Hamiltonian anyons can be simulated on NISQ quantum devices.

Our findings open several avenues for future research, \changes{both theoretical and practical. Promising platforms for implementation include quantum dots and ion traps, paving the way for direct manipulation of effective particle symmetry. On the theory side,} incorporating interactions could lead to even richer properties. In this initial approach, we penalized the space of nontrivial mixed symmetries, $\tilde{\cH}$, to achieve a clear delineation of symmetric and antisymmetric subspaces. Including this space in our analysis might increase the complexity of the problem and reveal new physical phenomena.

\begin{acknowledgments}
We thank Stefano Scali for advice on thermal state preparation in digital quantum computers.
We want to acknowledge funding from Ministry for Digital Transformation and of Civil Service of the Spanish Government through projects, PID2021-128970OA-I00 10.13039/501100011033 and QUANTUM ENIA project call - Quantum Spain project, and by the European Union through the Recovery, Transformation and Resilience Plan - NextGenerationEU within the framework of the Digital Spain 2026 Agenda, and also the FEDER/Junta de Andalucía program A.FQM.752.UGR20. JD acknowledges funding from the Engineering and Physical Sciences Research Council (EP/T518049/1). AT acknowledges funding from the Ministerio de Ciencia e Innovación of the Spanish Government through FPU20/02835.  MS acknowledges support from the Ministerio de Ciencia e Innovación of the Spanish Government under the Ramón y Cajal funding scheme RYC2021-032032-I as well as  FEDER funds C.EXP.256.UGR23 from the Junta de Andalucía. Finally, we are also grateful for the technical support provided by PROTEUS, the supercomputing center of the Institute Carlos I for Theoretical and Computational Physics in Granada, Spain.
	      
\end{acknowledgments}

\bibliography{refs}

\appendix
\section{Hilbert spaces for the $N$ anyons system}
\label{app:hilbert}
\subsection{Representation theory and Schur-Weyl Duality}
The following references provide all the information regarding the results presented in this section: \cite{serre1977linear,bishop1980tensor,fulton1991first,gustafson2003mathematical,takhtadzhian2008quantum,goodman2009symmetry,hall2013quantum,hall2013lie,woit2017quantum}. For simplicity and brevity, the proofs are omitted but can be found in the referenced sources. Note that, in this first subsection, we shall adopt $\C^d$ in place of a generic $d$-dimensional complex vector space $V$ without loss of generality, consistent with the notation of \cite{goodman2009symmetry}.

\subsubsection{Representations of $S_N$}

The most straightforward manner to compute and catalogue the irreducible representations of the permutation group of $N$ elements, $S_N$, is by means of the so-called \emph{Young tableaux}. This pictorial approach organizes the irreducible representations based on their behavior under permutations of two or more indices. The application to quantum mechanics becomes straightforward, enabling the distinction between symmetric spaces, antisymmetric spaces, and spaces with non-trivial symmetries.

\begin{definition}[Ordered, integer partitions]
    An ordered, integer partition, $\lambda$, of $N \in \mathbb{Z}^+$, is a $k$-tuple $\lambda = (\lambda_1, 
    \lambda_2, \ldots, \lambda_k)$, such that $ \lambda_1\geq\lambda_2\geq\ldots\geq\lambda_k$,  $\lambda_j\in\mathbb{Z}^+$ for all $1\leq j\leq k$, satisfying
    \begin{equation}
        \sum_{i=1}^k \lambda_i = N.
    \end{equation}
\end{definition}
There exists a particularly useful method of representing ordered integer partitions that allows for easy computation of their corresponding irreducible representations. The following two definitions introduce this representation known as \emph{Young diagrams} and \emph{Young tableaux}. 
\begin{definition}[Young diagram]
Let $\lambda$ be an integer partition of $N$.  A Young diagram corresponding $\lambda$ is an arrangement of $N$ boxes into $k$ rows such that 
the number of boxes in row $i\in\{1,\ldots,r\}$ is equal to $\lambda_i\in\lambda$.
\end{definition}
\begin{definition}[Young Tableau]
A Young tableau is a labeling of a Young diagram with the integers $(1,\ldots,N)$.  A canonical, or standard, Young tableau is one where the labeling is increasing from left to right, and from top to bottom.
\end{definition}

The permutation group of $N$ elements, $S_N$, acts naturally on any Young tableau by permuting its labels, providing a framework for representing the irreducible representations of $S_N$. Consider a Young tableau corresponding to the partition $\lambda$ of $N$, and define the following subgroups of $S_N$
	\begin{equation}
		\begin{split}
    			P_{\lambda} &: = \left\{\sigma \in S_N : \sigma \textrm{ preserves each row of } \lambda\right\},\\
    			Q_{\lambda} &: = \left\{\tau \in S_N : \tau \textrm{ preserves each column of } \lambda\right\}.
		\end{split}
	\label{app:subgroups}
	\end{equation}
To each of these two subgroups, the following two elements can be associated
	\begin{equation} 
    		\begin{split}
        		A_{\lambda} & = \sum_{\sigma \in P_{\lambda}} \sigma,\\
        		B_{\lambda} & = \sum_{\tau \in Q_{\lambda}} \textrm{sgn}(\tau) \tau,  
    		\end{split}
	\label{app:elements}
	\end{equation}
belonging to the group algebra of $S_N$---the complex vector space, $\C[S_N]$, with basis elements indexed by $S_N$. These elements enable the definition of the following construct.
\begin{definition}[Young Symmetrizer]
 The Young symmetrizer of the partition $\lambda$, $C_{\lambda}$, is the element of the associated group algebra $\C[S_N]$ defined as 
 	\begin{equation}
		C_{\lambda} :=A_{\lambda} B_{\lambda}.
	\label{app:Young_symmetrizer}
	\end{equation}
\end{definition}
 
The action of $S_N$ on $N$ copies of any $d$-dimensional complex vector space $\C^d$, $\bigotimes^N \C^d$, by way of permuting its indices, allows 
to define a natural group algebra representation $\C[S_N] \rightarrow \textrm{End}\left(\bigotimes^N \C^d\right)$. As we are interested in the symmetric and 
antisymmetric subspaces, the images of $A_{\lambda},B_{\lambda} \subset \C[S_N]$ into $\bigotimes^N \C^d$ are
	\begin{equation}
		\begin{split}
    			\textrm{Im}A_{\lambda} & =  \operatorname{Sym}^{\lambda_1} \C^d \otimes \ldots \otimes \operatorname{Sym}^{\lambda_k} \C^d  
			\subset \bigotimes^N \C^d,\\
			\textrm{Im}B_{\lambda} & = \operatorname{Alt}^{\lambda_1} \C^d \otimes \ldots \otimes \operatorname{Alt}^{\lambda_k} \C^d  
			\subset \bigotimes^N \C^d,
		\end{split}
	\label{app:images}
	\end{equation}
respectively, where $\operatorname{Sym}^{m}\C^d$ is the totally symmetric subspace of $\bigotimes^{m}\C^d$, and 
$\operatorname{Alt}^{m} \C^d$ is the totally antisymmetric subspace of $\bigotimes^{m}\C^d$, for $m \in \left\{\lambda_1, \ldots, \lambda_k\right\}$. The following theorem then establishes 
a connection between the Young symmetrizer and the irreducible representations of $S_N$ \cite{fulton1991first,goodman2009symmetry}.
\begin{theorem}
    The image of the Young symmetrizer $C_{\lambda}$ in $\C[S_N]$
	\begin{equation}
    		\mathrm{Im}C_{\lambda}= \C[S_N] C_{\lambda} \equiv \cV_{\lambda},
	\label{app:irrep}
	\end{equation}
    is an irreducible representation of the group $S_N$. Every irreducible representation of $S_N$ is isomorphic to $\cV_{\lambda}$ for some partition 
    $\lambda$.
\end{theorem}
The irreducible representation $\cV_{\lambda}$ is referred to as a \emph{Specht module}. 

By way of example, consider the Young tableaux consisting of a single row, $\lambda=(d)$, and a single column $\lambda = (1,1,\ldots, 1):=(\bf 1)$. 
For $\lambda = (d)$, $P_{(d)} = S_N$ and $Q_{(d)} = \{1\}$ so that
	\begin{equation}
    		C_{(d)} = A_{(d)} = \sum_{\sigma \in S_N} \sigma ,
	\label{app:symmetric}
	\end{equation} 
which implies that the image of $C_{(d)}$ on the vector space $\bigotimes^N \C^d$ is $\operatorname{Sym}^{d} \C^d$---the totally symmetric subspace of 
$\bigotimes^N \C^d$. Similarly, for $\lambda=(\bf 1)$ , $P_{(\bf 1)} = \{1\}$ and $Q_{(\bf 1)} = S_N$ so that
	\begin{equation}
    		C_{(\bf 1)} = B_{(\bf 1)} = \sum_{\tau\in S_N} \textrm{sgn}(\tau) \tau,
	\label{app:antisymmetric}
	\end{equation}
meaning that the image of $C_{(\bf 1)}$ on $\bigotimes^N \C^d$ is $\operatorname{Alt}^{d} \C^d$---the totally antisymmetric subspace of $\bigotimes^N \C^d$. The dimension of the irreducible representation $\cV_{\lambda}$ can be obtained via the Hook length formula
	\begin{equation}
    		\dim \cV_{\lambda} = \dfrac{N!}{\prod_{i,j} h(i,j)}:=m_{\lambda},
	\label{app:dimension}
	\end{equation}
where $h(i,j)$ is the hook length of the box located in the $i$th row and $j$-th column given by  
	\begin{equation}
		h(i,j)  = (\lambda_i-j)+(\lambda'_j-i)+1 ,
	\label{app:hooklength}
	\end{equation}
with $\lambda'_j$ the number of boxes in the $j$-th column (equivalently the number of boxes in the $j$-th row of the transpose Young diagram corresponding to 
$\lambda$). For $\lambda = {(\bf 1)}, (N)$, $\dim \cV_{(\bf 1)} = \dim \cV_{(N)} = 1$. This expression for the dimension of $\cV_{\lambda}$ allows 
us to determine the number of Young tableaux of a given Young diagram, i.e., $m_{\lambda}$ provides the number of tableaux for a given $\lambda$.

\subsubsection{Schur-Weyl duality}

In addition to the action of $S_N$, the vector space $\bigotimes^N \C^d$ naturally carries the action of $\textrm{GL}(d,\C)$. This is particularly relevant when
$\bigotimes^N \C^d$ corresponds to the state space of $N$ identical particles with a $d$-valued degree of freedom. This result by Schur and Weyl~\cite{schur1901klasse,schur1927rationalen,weyl1946classical} establishes a connection between the irreducible components of $S_N$ and those of $\mathrm{GL}(d,\mathbb{C})$, enabling the determination of the collective behavior of the degrees of freedom of particles associated with $\mathrm{GL}(d,\mathbb{C})$ under particle exchange~\cite{Fanizza2022}.  

Specifically, for all integers $d\geq 0$, let $\pi:\textrm{GL}(d,\C)\to\bigotimes^N \C^d$, and $T:S_N\to\bigotimes^N \C^d$ be representations 
of $\textrm{GL}(d,\C)$ and $S_N$ respectively whose action on $\bigotimes^N \C^d$ is defined as   
	\begin{equation}
		\begin{split}
   			\pi(g) (v_1 \otimes \ldots \otimes v_d) &= g (v_1) \otimes \ldots \otimes g(v_N), \\
		 	T(\sigma) (v_1 \otimes \ldots \otimes v_d) &=  v_{\sigma^{-1}(1)} \otimes \ldots \otimes v_{\sigma^{-1}(N)}, ,
		\end{split}
	\label{app:actions}
	\end{equation}
for $g\in \textrm{GL}(d,\C), \sigma\in S_N$, where $v_i\in\C^d, \textrm{ for } i=1,\ldots,N$. As it is possible to permute the factors without changing the action of $\textrm{GL}(d,\C)$, the actions of these two groups commute.  
In fact, the spans of the images of $\textrm{GL}(d,\C)$ and $S_N$ in $\textrm{End}\left(\bigotimes^N \C^d\right)$ act as double centralizers. This result 
establishes a profound connection between the representations of these two groups (for a more detailed discussion about mutual centralizers,  
see~\cite{goodman2009symmetry}).  To elucidate on this relation we require the following definition.
\begin{definition}[Schur functor]
The Schur functor $\mathbb{S}_{\lambda}$ of a partition $\lambda$ is the image of the Young symmetrizer $C_{\lambda}$ on $\bigotimes^N \C^d$.
Moreover, $\mathbb{S}_{\lambda}\C^d$ is a representation of $\emph{\textrm{GL}(N,\C)}$.
\end{definition}

The Schur functor satisfies the following properties.  Let $U,V,W \in \C^d$, and let $f: V \to W$, $g: U\to V$ be linear maps.  Then 
$\mathbb{S}_{\lambda}f : \mathbb{S}_{\lambda} V \to \mathbb{S}_{\lambda} W$ such that for any $\Psi\in\mathbb{S}_{\lambda} V$ 
it holds 
	\begin{equation}
		\mathbb{S}_{\bm\lambda}f(\Psi) = f^{\otimes N} \circ \Psi\in \mathbb{S}_{\lambda} W  .
	\label{app:functorality} 
	\end{equation}	
Moreover, $\mathbb{S}_{\lambda}(f \circ g) = \mathbb{S}_{\lambda}f \circ \mathbb{S}_{\lambda}g$ and $\mathbb{S}_{\lambda}(\mathrm{Id}_V) = \mathrm{Id}_{\mathbb{S}_{\lambda}}$. With this, we can now state the content of Schur-Weyl duality.
\begin{theorem}[Schur-Weyl duality]\label{theorem:schur_weyl}
The decomposition  
	\begin{equation}
    		\bigotimes^N \C^d \cong \bigoplus_{\lambda} \cV_{\lambda} \otimes \mathbb{S}_{\lambda} \C^d
	\label{app:SW1}
	\end{equation}
is a representation of $S_N \times \mathrm{\emph{GL}}(d,\C)$. Here, $\cV_{\lambda}$ ranges over all irreducible representations of $S_N$, \emph{\eqnref{app:irrep}}, and each 
$\mathbb{S}_{\lambda} \C^d$ is either an irreducible representation of $\mathrm{\emph{GL}}(d,\C)$ or zero. $\mathbb{S}_{\lambda} \C^d = 0$ if the partition of 
$N$ has more than $d$ parts, i.e., $\lambda > \dim \C^d$.
\end{theorem}

The proof of the Schur-Weyl theorem relies on the result stated in the \emph{double centralizer theorem}, see \cite{goodman2009symmetry}. See also Theorem 6.3, and Lemmas 6.22–6.23 in \cite{fulton1991first} for the proof.

As an example, consider the totally symmetric and antisymmetric subspaces, corresponding to the partitions $\lambda = (N), \lambda=(\bf 1)$, of the space $\bigotimes^N\C^d$.  The Schur functors $\mathbb{S}_{(N)} \C^d$ and $\mathbb{S}_{(\bf 1)} \C^d$ correspond to $\operatorname{Sym}^N \C^d$ and 
$\operatorname{Alt}^N\C^d$ respectively. Note that $\operatorname{Alt}^N \C^d$ vanishes if $\dim \C^d < N$. According to Schur-Weyl duality, stated in Theorem \ref{theorem:schur_weyl}, we can write 
the following representation of $\textrm{GL}(N,\C)$
	\begin{equation}
    		\bigotimes^N \C^d \cong \bigoplus_{\lambda} \left(\mathbb{S}_{\lambda} \C^d\right)^{\oplus m_{\lambda}}, 
	\label{app:SW2}
	\end{equation}
and the irreducible representations of $\textrm{GL}(N,\C)$ can be obtained directly from those of $S_N$. 
For instance, for $N=2$, \eqnref{app:SW2} reduces to 
	\begin{equation}
   		\C^d \otimes \C^d \cong  \operatorname{Sym}^2 \C^d \oplus \operatorname{Alt}^2 \C^d,
	\label{app:SW3}
	\end{equation}
which is the canonical decomposition of a product of two vector spaces into their symmetric and antisymmetric parts.  
\begin{figure}
    \centering
    \includegraphics[width=1\linewidth]{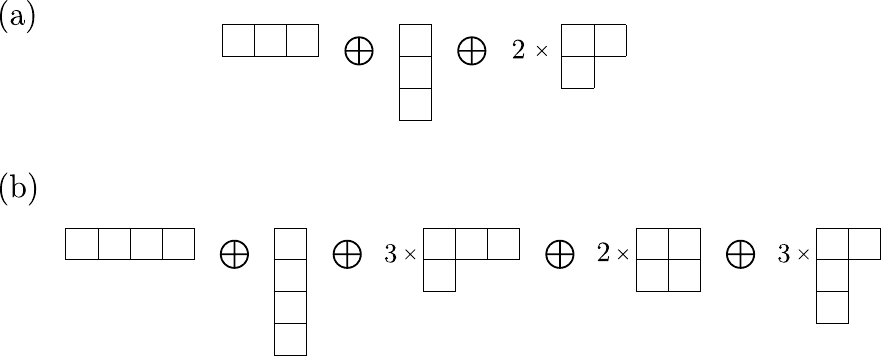}
    \caption{Diagrammatic representations of the (a) $N=3$ decomposition, see Eq. (\ref{app:c3}), and (b) $N=4$ decomposition, see Eq. (\ref{app:c4}).}
    \label{fig:examples_c3_c4}
\end{figure}
For $N=3, 4$, \eqnref{app:SW2} reads
	\begin{equation}\label{app:c3}
		\begin{split}
   			\bigotimes_{k=1}^3 \C^d_k & \cong \operatorname{Sym}^3 \C^d \oplus \left(\mathbb{S}_{(2,1)} \C^d\right)^{\oplus 2} \oplus \operatorname{Alt}^3 \C^d,
            \end{split}
            \end{equation}
            \begin{equation}\label{app:c4}
            \begin{split}
   			\bigotimes_{k=1}^4 \C^d_k & \cong \operatorname{Sym}^4 \C^d \oplus \operatorname{Alt}^4 \C^d\\
   			& \oplus \left(\mathbb{S}_{(3,1)} \C^d\right)^{\oplus 3} \oplus \left(\mathbb{S}_{(2,2)} \C^d\right)^{\oplus 2} \oplus \left(\mathbb{S}_{(2,1,1)} \C^d\right)^{\oplus 3}.
		\end{split}
	\end{equation}
The non-trivial irreducible representation $\mathbb{S}_{(2,1)} \C^d$ appears twice for the case of $N=3$, whilst the non-trivial irreducible representations 
$\mathbb{S}_{(3,1)} \C^d$,  $\mathbb{S}_{(2,2)} \C^d$, and $\mathbb{S}_{(2,1,1)} \C^d$ appear three times, twice, and three times respectively for the case of $N=4$. The diagrammatic representation of these examples is depicted in Fig. \ref{fig:examples_c3_c4}. As we will be solely interested in the  
totally symmetric and antisymmetric representations of $S_N$, we can write \eqnref{app:SW2} as
	\begin{equation}
    		\bigotimes^N \C^d \cong \operatorname{Sym}^N \C^d \oplus \operatorname{Alt}^N \C^d \oplus \mathscr{V} \textrm{(other)},
	\label{app:decomposition_general}
	\end{equation}
where $\mathscr{V} \textrm{(other)}$ accounts for all other \emph{exotic} symmetries that may appear in the decomposition. 

We conclude this section by computing the dimension of the vector spaces corresponding to the symmetric and antisymmetric subspaces of $\bigotimes^N\C^d$, as well as providing an orthonormal basis for each one.
\begin{proposition}
Let $\lambda$ be a tableau of dimension $N$ and $\dim \C^d = d$. The dimension of the Schur functor $\mathbb{S}_{\lambda} \C^d$ is
    \begin{equation}
        \dim \mathbb{S}_{\lambda} \C^d = \prod_{i,j} \dfrac{\lambda_i - \lambda_j + j -i}{j - i} = \dfrac{m_{\lambda}}{N!} \prod_{i,j} (d-i+j),
    \label{app:irrep_dimensions}
    \end{equation}
where the products are over the pairs $(i,j)$ that label the row and column of $\lambda$ respectively.
\end{proposition}
Analyzing the completely symmetric and antisymmetric cases, $m_{(N)} = m_{(\bf 1)} = 1$, as they are the only irreducible representations of the corresponding tableau. 
Then, the dimension of the symmetric and antisymmetric subspaces are directly obtained by computing the dimension of the Schur functor for these tableaux, i.e.,
	\begin{equation}
    		\begin{split}
        		\dim \mathbb{S}_{(N)} \C^d &  = \dfrac{\prod_{i=1, j=1,\ldots,N} (d-i+j)}{N!},\\
        		& = \binom{d + N -1}{N}\\
    			\dim \mathbb{S}_{(\bf 1)} \C^d &  = \dfrac{\prod_{i=1,\ldots,N, j=1} (d-i+j)}{N!}\\
        		& = \binom{d}{N}.
		\end{split}
    	\label{eqn:dim_sym_anti}
	\end{equation}
A basis for these spaces can be constructed using the following proposition \cite{bishop1980tensor}.
\begin{proposition} \label{prop:basis}
 Let $\{e_i\}_{i=1}^d$ a basis of $\C^d$. For the totally symmetric and antisymmetric subspaces, we can construct the following orthogonal basis
	\begin{equation}
        	\begin{split}
            		B_{\operatorname{Sym}^N  \C^d} & = \left\{ \sum_{\sigma \in S_N} e_{\sigma(1)} \otimes \ldots \otimes e_{\sigma(N)}  \right\},\\
            		B_{\operatorname{Alt}^N  \C^d} & = \left\{ \sum_{\tau \in S_N}  \mathrm{\emph{sgn}}(\tau) e_{\tau(1)} \otimes \ldots \otimes e_{\tau(N)}  \right\}.
        	\end{split}
    	\end{equation}
\end{proposition}
In the context of quantum mechanics, the dimensions of these spaces are directly related to the degrees of freedom available to $N$ identical particles with $d$ 
internal states. For bosonic systems (totally symmetric), $\dim \mathbb{S}_{(N)} \C^d$ corresponds to the number of possible configurations of $N$ bosons occupying 
$d$ states. In the case of fermionic systems (totally antisymmetric), $\dim \mathbb{S}_{(\bf 1)} \C^d $ represents the number of possible configurations for $N$ 
fermions obeying the Pauli exclusion principle \footnote{The determination of the dimension of the spaces and their bases is a straightforward derivation in elementary tensor algebra. Here we show that the same result can be derived from a purely representation theory approach.}.

\subsection{Hilbert spaces}\label{app:subsec_hilbert}
Consider a quantum particle with spin $j$. In general, the total Hilbert space describing it has two terms
\begin{equation}\label{eqn:hilbert_1p}
    \cH = \mathscr{H} \otimes V_{d},
\end{equation}
where $\mathscr{H}$ is the space related to the internal degrees of freedom, e.g. the position of the particle, and $V_{d}$ is a complex vector space of dimension $d$, with $d=2j+1$, carrying an irreducible representation of $\mathrm{SU}(d)$. It is convenient to represent the total state of the system as $\ket{\Psi} \in \cH$, which can be written as $\Psi = \psi (\boldsymbol{r}) \chi(s)$, where $\psi (\boldsymbol{r}) \in \mathscr{H}$ with $\boldsymbol{r} \in \R^3$, and $\chi(s) \in V_d$. This variable $s$ depends on the spin of the particle, and takes the $2j+1$ values from $\{-j,-j+1,\ldots,j+1,j\}$. The vector space $V_d$ is isomorphic to the complex numbers so that $V_d \cong \C^{d} = \C^{2j+1}$. Note that we are abusing the notation when writing $\chi(s)$. What we are really doing is identifying the complex space $\mathbb{C}^{2j+1}$ with the space of functions
\begin{equation} 
\chi: \left\{-j,-j+1,\ldots,j-1,j \right\} \longrightarrow \mathbb{C}, 
\end{equation}
so that, for every element $e_i \in B_{\mathbb{C}^{2j+1}} = \left\{e_1,\ldots,e_{2j+1}\right\}$, we assign the function $\chi_i$ acting as
\begin{equation} 
\chi_i (j+1-i) = 1; \ 0 \textrm{ otherwise}, \end{equation}
for all $i = 1,\ldots,2j+1$. This identification generalizes the concept of spinor to particles with any arbitrary spin, see \cite{takhtadzhian2008quantum} for further details.

Consider a quantum system composed of $N$ identical particles with spin $j$. The total Hilbert space corresponding to the total system is $N$ times the individual one, Eq. (\ref{eqn:hilbert_1p}), namely
\begin{equation}\label{eqn:hilbert_Np}
   \cH^{\otimes N} = \mathscr{H}^{\otimes N}\otimes V_d^{\otimes N}.
\end{equation}
As stated in Eq. (\ref{app:actions}), there is a natural action of the permutation group of $N$ elements, $S_N$, on the total Hilbert space given by Eq. (\ref{eqn:hilbert_Np}). Consider the general $N$-particle Hamiltonian acting on the system
\begin{equation}\label{eqn:total_ham}
    H_N = -\dfrac{\hbar}{2m}\sum_{i=1}^N p_i^2 + \sum_{i=1}^N \phi_{\mathrm{int}}(\boldsymbol{r}_i) + \sum_{i<j} \phi_{\mathrm{ext}}\left(\boldsymbol{r}_i - \boldsymbol{r}_j\right),
\end{equation}
where $m$ is the mass of the particles, which has been assumed to be equal for all of them without loss of generality. Then, by virtue of this natural action of the permutation group, $[H_N, T_N(\sigma)] = 0$, for any given $\sigma \in S_N$, and any representation $T_N$ of $S_N$ on the total Hilbert space. The direct consequence of this result dictates that $H_N$ can be restricted to invariant subspaces of $\cH^{\otimes N}$, and particularly to $\cH^{(\mathrm{Sym})}$ and $\cH^{(\mathrm{Alt})}$. That is, for bosonic particles, the Hilbert space for $N$-particle systems is the totally symmetric subspace $\cH^{(\mathrm{Sym})}$ of the $N$-fold product in Eq. (\ref{eqn:hilbert_Np}). The same applies for fermions with the totally antisymmetric subspace $\cH^{(\mathrm{Alt})}$. Note that the total Hamiltonian, Eq. (\ref{eqn:total_ham}), does not act on the spin variables.

As a consequence of the Schur-Weyl duality, Eqs. (\ref{app:SW1})-(\ref{app:decomposition_general}), the total Hilbert $\cH^{\otimes N}$ space decomposes into a direct sum of irreducible subspaces $\cH^{(\lambda)}$, that correspond to the different symmetries of the $N$ particles under permutations. This means that the irreducible representations of $S_N$ are connected to the ones of $\mathrm{SU}(d)$ which are carried by the spaces $V_d$, allowing us to elucidate the total spin of the corresponding wavefunctions. Then,
    \begin{equation}\label{app:hilbert_decomposition}
         \cH^{\otimes N} \cong \bigoplus_{\lambda} \cH^{(\lambda)},
    \end{equation}
where the terms $\cH^{(\lambda)}$ are the subspaces carrying the irreducible representations of $S_N$ labeled by $\lambda$. The totally symmetric and antisymmetric subspaces decompose as 
    \begin{equation}\label{app:total_hilbert_sym}
    \begin{split}
        \cH^{(\mathrm{Sym})}&= \left[\mathscr{H}^{\otimes N}\right]^{(\mathrm{Sym})}\otimes \left[V_d^{\otimes N}\right]^{(\mathrm{Sym})} \\
        &\oplus \left[\mathscr{H}^{\otimes N}\right]^{(\mathrm{Alt})}\otimes \left[V_d^{\otimes N}\right]^{(\mathrm{Alt})} \\
        &\oplus \tilde{\cH}^{(\mathrm{Sym})},
    \end{split}
    \end{equation}
\begin{equation}\label{app:total_hilbert_alt}
    \begin{split}
        \cH^{(\mathrm{Alt})}&= \left[\mathscr{H}^{\otimes N}\right]^{(\mathrm{Sym})}\otimes \left[V_d^{\otimes N}\right]^{(\mathrm{Alt})} \\
        &\oplus \left[\mathscr{H}^{\otimes N}\right]^{(\mathrm{Alt})}\otimes \left[V_d^{\otimes N}\right]^{(\mathrm{Sym})} \\
        &\oplus \tilde{\cH}^{(\mathrm{Alt})},
    \end{split}
    \end{equation}
where $\tilde{\cH}^{(\mathrm{Sym})}, \tilde{\cH}^{(\mathrm{Alt})}$ correspond to the product spaces of non-trivial symmetries which, when multiplied, give rise to totally symmetric and antisymmetric spaces respectively. In the particular case of two particles with spin $1/2$, the total Hilbert space is directly
\begin{equation}\label{app:hilbert_2p}
    \cH^{\otimes 2} = \mathscr{H}^{\otimes 2} \otimes V_2^{\otimes 2}.
\end{equation}  
Since the system consists of two fermions, we are interested in the totally antisymmetric subspace, which, from Eq.~(\ref{app:total_hilbert_alt}), decomposes as  
\begin{equation}\label{app:hilbert_2p_alt}
    \begin{split}
        \cH^{(\mathrm{Alt})} &= \left[\mathscr{H}^{\otimes 2}\right]^{(\mathrm{Sym})} \otimes \left[V_2^{\otimes 2}\right]^{(\mathrm{Alt})} \\
        &\quad \oplus \left[\mathscr{H}^{\otimes 2}\right]^{(\mathrm{Alt})} \otimes \left[V_2^{\otimes 2}\right]^{(\mathrm{Sym})},
    \end{split}
\end{equation}  
where $V_2^{\otimes 2} \cong \mathbb{C}^2 \otimes \mathbb{C}^2$. Using the general expressions for the basis in the totally symmetric and antisymmetric spaces in Proposition \ref{prop:basis}, the basis for this particular case are
\begin{equation}\label{app:states:2_part_sym}
    \operatorname{Sym}^2 \mathbb{C}^2 = \textrm{span}\left\{ e_1 \otimes e_1,  e_1 \otimes e_2 + e_2 \otimes e_1,  e_2 \otimes e_2 \right\},
\end{equation}  
\begin{equation}\label{app:states:2_part_alt}
    \operatorname{Alt}^2 \mathbb{C}^2 = \textrm{span}\left\{ e_1 \otimes e_2 - e_2 \otimes e_1 \right\},
\end{equation}  
where $\{e_1, e_2\}$ is a basis of $\mathbb{C}^2$. This particular case is straightforward to compute and can be directly derived from the irreducible representations of $\mathrm{SU}(2)$ using the Clebsch-Gordan decomposition. Let us label the vector space $V$ by the value of the associated spin, $j$ and $j'$. The product of two spaces $V_j$ and $V_{j'}$ is then decomposed as  
\begin{equation}
    V_j \otimes V_{j'} = \bigoplus_{l = |j-j'|}^{|j+j'|} V_l.
\end{equation}  
In our specific case, where $j = j' = 1/2$, this results in  
\begin{equation}
    V_{1/2} \otimes V_{1/2} = V_0 \oplus V_1.
\end{equation}  
The basis for $V_0$ can be computed directly, yielding the state in Eq. (\ref{app:states:2_part_alt}), while the basis for $V_1$ corresponds to the state in Eq. (\ref{app:states:2_part_sym}). The dimension for $V_0$ is one, and for $V_1$ is three, directly derived from the relation between the dimension and spin, $l=2j+1$. This is the reason why in physics terminology one refers to the antisymmetric space as \emph{singlet}, and to the symmetric one as \emph{triplet}. For more details, consult \cite{hall2013lie,woit2017quantum}. Although the Clebsch-Gordan decomposition is manageable for two particles, it becomes increasingly complex when applied to products involving more terms. For systems with more than a few particles, the use of Schur-Weyl duality becomes essential to handle the computations effectively.

\section{Calculations of thermodynamic quantities}\label{app:thermodynamics}

In this appendix, we expand on some of the calculations presented in Sec. \ref{sec:thermodynamics}, concerning the thermal properties of Hamiltonian anyons in a harmonic potential. For completeness, we first note that the $N$-particle, noninteracting, one-dimensional harmonic oscillator Hamiltonian is explicitly given (before symmetrization) by
\begin{equation}
\begin{split}
    H_\mathrm{HO} & = \sum_{k=1}^N \left\{ \id_{\mathscr{H}}^{\otimes (k-1)} \otimes \left[\sum_{n=0}^\infty \hbar \omega \left( n + \frac{1}{2} \right) \ket{n}\bra{n}  \right] \right.\\
    & \left.  \otimes \ \id_{\mathscr{H}}^{\otimes (N-k)} \right\}.
    \end{split}
\end{equation}
where $\ket{n}$ denotes the state where the $k$-th particle occupies the $n^\mathrm{th}$ energy level. The task of computing the canonical partition function for bosons or fermions with this Hamiltonian is nontrivial, but can be achieved by using the grand canonical ensemble to obtain a recursion relation for successive numbers of particles, $N$. See Ref. \cite{mullin_02} for more details.

To obtain the reduced thermal state of the harmonic oscillator degrees of freedom, we proceed from Eq. \eqref{eqn:ham_exponentiate}
\begin{equation}
    \begin{split}
        \tau_{\mathscr{H}^{\otimes N}} &= \Tr_{\left(\mathbb{C}^d \right)^{\otimes N}} \left\{\left[\frac{1}{Z} e^{-\beta H}\right] ^\mathrm{(Alt)} \right\}\\
        &= \frac{1}{Z} \left[ e^{-\beta H_\mathrm{HO}} \right]^\mathrm{(Alt)} \Tr\left[\Pi_{\left(\mathbb{C}^d \right)^{\otimes N}}^\mathrm{(Sym)}\right]\\
        & + \frac{1}{Z} e^{-\beta\nu}\left[ e^{-\beta H_\mathrm{HO}} \right]^\mathrm{(Sym)} \Tr\left[\Pi_{\left(\mathbb{C}^d \right)^{\otimes N}}^\mathrm{(Alt)}\right]\\
        &= \binom{d+N-1}{N}\frac{Z_\mathrm{F}}{Z}\left(\frac{1}{Z_\mathrm{F}} \left[ e^{-\beta H_\mathrm{HO}} \right]^\mathrm{(Alt)}\right)\\
        &  + \binom{d}{N} e^{-\beta\nu}\frac{Z_\mathrm{B}}{Z}\left(\frac{1}{Z_\mathrm{B}} \left[ e^{-\beta H_\mathrm{HO}} \right]^\mathrm{(Sym)}\right)\\
        & = \binom{d+N-1}{N}\frac{Z_\mathrm{F}}{Z} \tau_\mathrm{F} + \binom{d}{N} e^{-\beta\nu}\frac{Z_\mathrm{B}}{Z} \tau_\mathrm{B},
    \end{split}
\end{equation}
where the fermionic/bosonic thermal states are $\tau_\mathrm{F/B} = \frac{1}{Z_\mathrm{F/B}}\left[e^{-\beta H_\mathrm{HO}}\right]^\mathrm{(Alt/Sym)}$. The binomial factors appear directly when tracing out the projectors $\Pi_{\left(\mathbb{C}^d \right)^{\otimes N}}^\mathrm{(Sym/Alt)}$, providing the dimension of the symmetric/antisymmetric auxiliary subspaces as in Eq. \eqref{eqn:dim_sym_anti}. 

We define $p_\mathrm{F}$ as the prefactor on $\tau_\mathrm{F}$ 
\begin{equation}
        p_\mathrm{F} = \binom{d+N-1}{N}\frac{Z_\mathrm{F}}{Z}
        = \frac{\binom{d+N-1}{N}Z_\mathrm{F}}{ \binom{d+N-1}{N}Z_\mathrm{F} + \binom{d}{N} e^{-\beta\nu}Z_B }.
\end{equation}
Note that, from Eq. \eqref{eqn:Z_n_fermion_boson}, $\frac{Z_B}{Z_F} = e^{\frac{1}{2}\beta\hbar\omega(N^2-N)}$. Let us define $h(d,N) := \ln\binom{d+N-1}{N} - \ln\binom{d}{N}$. Then, $p_\mathrm{F}$ can be trivially expressed as in Eq. \eqref{eqn:pf}. 

The internal energy can be computed from the partition function as follows
\begin{equation}
    \begin{split}\label{app:internal_energy}
        U 
        &= - \frac{1}{Z} \frac{\partial Z}{\partial\beta}\\
        &= - \frac{1}{Z} \left[ \binom{d+N-1}{N} \frac{\partial Z_\mathrm{F}}{\partial\beta} + \binom{d}{N} \frac{\partial}{\partial\beta} \left(e^{-\beta\nu} Z_\mathrm{B}\right)\right]\\
        &= \binom{d+N-1}{N} \frac{Z_\mathrm{F}}{Z} U_\mathrm{F} + \binom{d}{N} e^{-\beta\nu} \frac{Z_\mathrm{B}}{Z} (\nu + U_\mathrm{B})\\
        &= p_\mathrm{F} U_\mathrm{F} + (1-p_\mathrm{F})(\nu + U_\mathrm{B}),
    \end{split}
\end{equation}
where the fermionic and bosonic internal energies can be straightforwardly calculated from Eq. \eqref{eqn:Z_n_fermion_boson} using that $U_\mathrm{F/B} = - \frac{\partial}{\partial\beta} \ln Z_\mathrm{F/B}$.

\section{Capacities and Phase Transitions}
\label{app:heat_capacity}
In this appendix, we compute the first and second derivatives of the internal energy for $N$ Hamiltonian anyons, Eq. (\ref{app:internal_energy}), to identify first- and second-order phase transitions. The first derivative, which corresponds to the heat capacity with respect to each of the parameters $X\in\{T,\nu,\omega\}$ of the system are given by   
\begin{equation}\label{app:heat_capacity_general}
    \begin{split}
    C_X\equiv \UX &= \pf \left(\UFX-\UBX -\nuX\right)+\UBX +\nuX\\
    &+\pfX\left(\UF-\UB-\nu\right)\\
    &= \pf \left(\UFX-\UBX -\nuX\right)  +\UBX+\nuX\\
    &-\pf^2\left(\phiX\right)e^\phi\left(\UF-\UB-\nu\right),
    \end{split}
\end{equation}
where the parameter $\phi$ reads
\begin{equation}
\phi = \frac{\hbar\omega\beta}{2}N(N-1)-\nu\beta-h(d,N).
\label{app:phi}
\end{equation}
To derive \eqnref{app:heat_capacity_general}, it is convenient to make use of the following identity for $\pf$
    \begin{equation}
            \pfX = \frac{\partial}{\partial X}\left(\frac{1}{1+e^\phi}\right)= -\pf^2\left(\phiX\right)e^\phi.
    \label{app:derspf}
    \end{equation}
Noting that neither $\UB$ nor $\UF$ depend explicitly on $\nu$, i.e.,
\begin{equation}
    \UBnu = \UFnu = 0,
\end{equation}
and that their derivatives with respect to  $\beta$ and $\omega$ are, respectively

\begin{equation}\label{app:dersinternalU}
\begin{split}
    \UBbeta &= \UFbeta = -
    \frac{\hbar^2\omega^2}{4}\sum_{k=1}^{N} k^2 \csch^2\left(\frac{k\hbar\beta\omega}{2}\right),\\
        \UBomega &= \frac{\hbar N}{2} - \hbar\sum_{k=1}^{N} \frac{k}{(1 - e^{k\beta\hbar\omega})^2} \\
        &- \frac{\hbar^2\beta\omega}{4} \sum_{k=1}^{N} k^2 \csch^2\left(\frac{k\hbar\beta\omega}{2}\right),\\
        \UFomega &= \frac{\hbar N^2}{2} - \hbar\sum_{k=1}^{N} \frac{k}{(1 - e^{k\beta\hbar\omega})^2} \\
        & - \frac{\hbar^2\beta\omega}{4} \sum_{k=1}^{N} k^2 \csch^2\left(\frac{k\hbar\beta\omega}{2}\right).
\end{split}
\end{equation}
From these derivatives, we can compute the heat capacities, Eq. (\ref{app:heat_capacity_general})
\begin{equation}\label{app:heat_capacities}
    \begin{split}
        \frac{\partial U}{\partial T}&=k_{\textrm{B}}\beta^2\left\{\pf^2e^\phi\left[\frac{\hbar\omega}{2}N(N-1)-\nu\right]^2 -\UBbeta\right\},\\
        \Unu&= 1-\pf +\pf^2\beta e^\phi\left[\frac{\hbar\omega}{2}N(N-1)-\nu\right],\\
        \Uomega&= \frac{\hbar \pf}{2}N(N-1)+\UBomega\\
        &-\frac{\hbar\beta \pf^2 e^\phi}{2}N(N-1)\left[\frac{\hbar\omega}{2}N(N-1)-\nu\right].
    \end{split}
\end{equation}
Having computed the derivatives, we seek to identify non-analytical behavior for the heat capacities in the asymptotic limit of a large number of particles. Specifically, we are interested in the heat capacity per particle, but as we have mixtures of both bosons and fermions---with 
the latter obeying the Pauli exclusion principle---we search for non-analyticity of the following limit 
\begin{equation}\label{appa:limit}
    \lim_{N\to\infty}\frac{1}{N^2}\UX .
\end{equation}
Noting that the derivatives of the bosonic internal energies vanishes in this limit 
\begin{equation}
\begin{split}
\lim_{N\to\infty}\frac{1}{N^2}\UBbeta&=0\\
\lim_{N\to\infty}\frac{1}{N^2}\UBomega&=0, 
\end{split}
\end{equation}
the limits for the total internal energy with respect to the parameters $X\in\{T,\nu,\omega\}$, see Eq. (\ref{app:heat_capacity_general}), are dominated by the fermionic behavior in the following form 

\begin{equation*}\label{app:Capacities}
\begin{split}
        &\lim_{N\to\infty}\frac{1}{N^2}\frac{\partial U}{\partial T}\\
        &\hspace{2em}= k_{\textrm{B}}\beta^2 \lim_{N\to\infty}\left\lbrace
        \frac{\pf^2 e^\phi}{N^2}\left[\frac{\hbar\omega}{2}N(N-1)-\nu\right]^2\right\rbrace,
\end{split}
\end{equation*}
\begin{equation}
\begin{split}
        &\lim_{N\to\infty}\frac{1}{N^2}\Unu \\
        &\hspace{2em} = \lim_{N\to\infty}\left\lbrace \frac{\beta \pf^2 
        e^\phi}{N^2}\left[\frac{\hbar\omega}{2}N(N-1)-\nu\right]\right\rbrace,
\end{split}
\end{equation}
\begin{equation*}
\begin{split}
        &\lim_{N\to\infty}\frac{1}{N^2}\Uomega \\
        & \hspace{2em} = \frac{\hbar \pf}{2} - \frac{\hbar\beta}
        {2}\lim_{N\to\infty}\left\lbrace \left[\frac{\hbar\omega}{2}N(N-1)-\nu\right]\right\rbrace.
\end{split}
\end{equation*}

Observe, however, that the parameters $\beta,\nu,\omega$ are not independent as is evident from \eqnref{app:phi}.  Importantly, note that $\pf$ experiences a sharp transition around the
value $\phi=0$, where its second derivative
    \begin{equation} \label{app:second_der_pf}
        \pfXX = \pf^2\left(\phiX\right)^2 e^\phi\left(2\pf e^\phi-1\right),
    \end{equation}
equals zero. Therefore, substituting
\begin{equation}
\nu = \frac{\hbar\omega}{2} N(N-1) -\frac{h(d,N)}{\beta} - \frac{\varepsilon}{\beta},
\label{app:smallepsilon}
\end{equation}
for $\varepsilon > 0$ into \eqnref{app:heat_capacities}, and taking the asymptotic limit, we obtain the following behavior for the capacities
\begin{equation*}
    \begin{aligned}
        \lim_{N\to\infty}\frac{1}{N^2}\frac{\partial U}{\partial T} &= k_{\textrm{B}}\beta^2 \lim_{N\to\infty}\left\lbrace 
        \frac{\pf^2 e^\phi}{\beta^2}\frac{[h(d,N)+\varepsilon]^2}{N^2}\right\rbrace\\
        &=\frac{k_{\textrm{B}}}{4},
    \end{aligned}
\end{equation*}

\begin{equation}\label{app:Capacities_epsilon}
    \begin{aligned}
        \lim_{N\to\infty}\frac{1}{N^2}\Unu &= \pf^2 
    e^\phi\lim_{N\to\infty}\left\lbrace\frac{h(d,N)+\varepsilon}{N^2}\right\rbrace = 0,
    \end{aligned}
\end{equation}

\begin{equation*}
    \begin{aligned}
        \lim_{N\to\infty}\frac{1}{N^2}\Uomega &=\frac{\hbar \pf}{2} - \frac{\hbar}
        {2}\lim_{N\to\infty}\left\lbrace h(d,N)+\varepsilon\right\rbrace= -\infty.
    \end{aligned}
\end{equation*}
Note that, in the thermodynamic limit $(N\to \infty)$, the function $h(d,N)$ tends to
\begin{equation} \label{app:limith}
     h(d,N)\to \begin{cases}
            \log\frac{N}{d} & \, \mathrm{ for } \,d \gg N,\\
            2N & \, \mathrm{ for } \, d=N .
            \end{cases}
\end{equation}
The divergence in the capacity with respect to $\omega$ occurs due to the fact that the ground state energy for Hamiltonian anyons jumps from linear to quadratic in the number of particles as $\omega$ approaches the value for which $\phi=0$.  This can be clearly seen in \figref{fig:anyonicity}(b), where the spread of the phase transition with respect to each of the three parameters is also illustrated. Finally, the second-order derivative of the internal energy is
\begin{widetext}
\begin{equation}
\begin{split}
    \UXX &= \pf^2\left(\frac{\partial\phi}{\partial X}\right) e^\phi\left\{\left(\frac{\partial\phi}{\partial X}\right)(2\pf e^\phi-1)\left[\frac{\hbar\omega}{2}N(N-1)-\nu\right]
    -2\left(\frac{\partial U_{\textrm{F}}}{\partial X}-\frac{\partial U_{\textrm{B}}}{\partial X}-\frac{\partial\nu}{\partial X}\right)\right\} \\
    &+ \pf\left(\frac{\partial^2 U_{\textrm{F}}}{\partial X^2}-\frac{\partial^2 U_{\textrm{B}}}{\partial X^2}\right)+\frac{\partial^2 U_{\textrm{B}}}{\partial X^2}  .
\end{split}
\end{equation}
\end{widetext}
Noting that the following relations for the second derivative of the bosonic and fermionic internal energies hold 
\begin{equation}
        \UFXX = \UBXX; \ \lim_{N\to\infty}\frac{1}{N^2}\UBXX = 0,
\end{equation}
and using \eqnref{app:smallepsilon}, the second derivatives of the internal energy read explicitly 
\begin{equation*}
    \begin{split}
        \frac{1}{N^2}\frac{\partial^2 U}{\partial T^2} & =\frac{\varepsilon k_\mathrm{B}}{8T}\frac{\left[h(d,N)+\varepsilon\right]^3}{N^2}+\frac{1}{N^2}\UBbbeta,
\end{split}
\end{equation*}
\begin{equation}\label{app:second_derivatives}
\begin{split}
        \frac{1}{N^2}\Unnu& = \frac{\varepsilon\beta}{8}\left[\frac{h(d,N)+\varepsilon}{N^2}\right]-\frac{\beta}{2N^2},
\end{split}
\end{equation}
\begin{equation*}
\begin{split}
        \frac{1}{N^2}\Uoomega & = \frac{\varepsilon\hbar}{16}\frac{N(N-1)}{N^2}\left[h(d,N)+\varepsilon\right] \\
         &-\frac{\hbar^2\beta}{8}(N-1)^2 + \frac{1}{N^2}\UBbbeta .
    \end{split}
\end{equation*}
In the asymptotic limit, the derivatives in Eq. (\ref{app:second_derivatives}) result in 
\begin{equation*}
    \begin{split}
        \lim_{N\to\infty}\frac{1}{N^2}\frac{\partial^2 U}{\partial T^2}&= \lim_{N\to\infty}\frac{\varepsilon k_\mathrm{B}}{8T} N = \infty,
    \end{split}
\end{equation*}
\begin{equation}
\begin{split}
        \lim_{N\to\infty}\frac{1}{N^2}\Unnu &= 0,
    \end{split}
\end{equation}
\begin{equation*}
\begin{split}
        \lim_{N\to\infty}\frac{1}{N^2}\Uoomega &= \lim_{N\to\infty} -\frac{\hbar^2\beta}{8} N^2=-\infty,
    \end{split}
\end{equation*}
exhibiting phase transitions both with respect to temperature and frequency.

\section{Further aspects of the work extraction cycles}\label{app:engines}

We here back up a few claims relating to the fermionizing/bosonizing Stirling cycle, and later the Otto cycle. 

\subsection{Stirling cycle: conditions for positive work extraction}

We here examine conditions under which the Stirling cycle extracts positive work. Our argument centers around derivatives of the free energy $F = -\frac{1}{\beta}\ln Z$. Considering first the derivative with respect to $\nu$, we have
\begin{equation}\label{nu_partial}
    \begin{split}
        \frac{\partial F}{\partial\nu} &= - \frac{\partial}{\partial\nu} \frac{1}{\beta}\ln Z
        = - \frac{1}{\beta} \frac{\partial}{\partial\nu}\ln\left[\frac{\binom{d+N-1}{N} Z_\mathrm{F}}{p_\mathrm{F}}\right]\\
        &= \frac{1}{\beta} \frac{\partial}{\partial\nu}\ln p_\mathrm{F}= \frac{1}{\beta} \frac{1}{p_\mathrm{F}} \frac{\partial p_\mathrm{F}}{\partial\nu}.
    \end{split}
\end{equation}
In the second line, we use that $Z = \frac{\binom{d+N-1}{N} Z_\mathrm{F}}{p_\mathrm{F}}$, which follows from the definition of $p_\mathrm{F}$ in Eq. \eqref{eqn:pf}. In the third, we used the fact that $\binom{d+N-1}{N} Z_\mathrm{F}$ does not depend on $\nu$. We now note that, from Eq. (\ref{app:derspf}) for $X=\nu$
\begin{equation}
    \begin{split}
        \frac{\partial p_\mathrm{F}}{\partial\nu}  &= \frac{\partial }{\partial\nu}\left\{1+ \exp[\frac{1}{2}N(N{-}1)\beta\hbar\omega - \beta\nu - h(d,N)]\right\}^{-1}\\
        &= p_\mathrm{F}^2  \beta\exp[\frac{1}{2}N(N{-}1)\beta\hbar\omega - \beta\nu - h(d,N)],
    \end{split}
\end{equation}
and therefore 
\begin{equation}
    \begin{split}
        \frac{\partial F}{\partial\nu} = p_\mathrm{F}\exp[\frac{1}{2}N(N{-}1)\beta\hbar\omega - \beta\nu - h(d,N)],
    \end{split}
\end{equation}
which is non-negative. Physically, $ - \frac{\partial F}{\partial\nu}$ represents the rate at which work is done by the system when raising $\nu$. We can therefore infer that raising $\nu$ always costs work, while lowering $\nu$ results in work extraction. However, the sign of the net work output of the whole cycle is still not clear. We consider the derivative $\frac{\partial^2 F}{\partial\beta\partial\nu}$
\begin{equation}
    \begin{split}
        \frac{\partial^2 F}{\partial\beta\partial\nu} &= p_\mathrm{F} \frac{\partial}{\partial\beta}\exp[\frac{1}{2}N(N{-}1)\beta\hbar\omega - \beta\nu - h(d,N)]\\
        &+ \exp[\frac{1}{2}N(N{-}1)\beta\hbar\omega - \beta\nu - h(d,N)] \frac{\partial p_\mathrm{F}}{\partial\beta} \\
        &= \exp[\frac{1}{2}N(N{-}1)\beta\hbar\omega - \beta\nu - h(d,N)] \\
        & \hspace{1em} \left\{ \left[\frac{1}{2}N(N{-}1)\hbar\omega - \nu\right]p_\mathrm{F} + \frac{\partial p_\mathrm{F}}{\partial\beta} \right\}.
    \end{split}
\end{equation}
Here, the derivative of $p_\mathrm{F}$ with respect to $\beta$, from Eq. (\ref{app:derspf}) for $X=\beta$, reads
\begin{equation}
    \begin{split}
        \frac{\partial p_\mathrm{F}}{\partial\beta}  &= \frac{\partial }{\partial\beta}\left\{1+ \exp[\frac{1}{2}N(N{-}1)\beta\hbar\omega - \beta\nu - h(d,N)]\right\}^{-1}\\
        &= -p_\mathrm{F}^2 \frac{\partial }{\partial\beta} \exp[\frac{1}{2}N(N{-}1)\beta\hbar\omega - \beta\nu - h(d,N)]\\
        &= -p_\mathrm{F}^2  \left[\frac{1}{2}N(N{-}1)\hbar\omega - \nu\right] \\
        &\hspace{1em} \exp[\frac{1}{2}N(N{-}1)\beta\hbar\omega - \beta\nu - h(d,N)],
    \end{split}
\end{equation}
so that
\begin{equation}
    \begin{split}
        \frac{\partial^2 F}{\partial\beta\partial\nu} &= \exp[\frac{1}{2}N(N{-}1)\beta\hbar\omega - \beta\nu - h(d,N)] \\
        &\hspace{1em}  \left[\frac{1}{2}N(N{-}1)\hbar\omega - \nu\right]\left(p_\mathrm{F} - p_\mathrm{F}^2\right).
    \end{split}
\end{equation}
In this expression, the exponential factor as well as $\left(p_\mathrm{F} - p_\mathrm{F}^2\right)$ are guaranteed to be non-negative, since $0\leq p_\mathrm{F}\leq 1$. Therefore, $\frac{\partial^2 F}{\partial\beta\partial\nu}$ takes the same sign as the remaining factor $\frac{1}{2}N(N{-}1)\hbar\omega - \nu$. In particular, this means that if $\nu < \frac{1}{2}N(N{-}1)\hbar\omega$, then $\frac{\partial^2 F}{\partial\beta\partial\nu}\geq 0$ and the work cost $\frac{\partial F}{\partial\nu}\delta\nu$ of raising through a small increment $\delta\nu$ is monotone-increasing with $\beta$. It follows that if $\beta_\mathrm{H}>\beta_\mathrm{C}$ and $\nu_2<\nu_1 < \frac{1}{2}N(N{-}1)\hbar\omega$, then the work cost of raising from $\nu_2$ to $\nu_1$ at $\beta_\mathrm{C}$ (i.e. fermionizing at low temperature) is less than the work \emph{extraction} when lowering from $\nu_1$ to $\nu_2$ at $\beta_\mathrm{H}$ (bosonizing at high temperature). The net result is positive work extraction. 

A reversal takes place if $\nu_2 > \nu_1 > \frac{1}{2}N(N{-}1)\hbar\omega$: the magnitude of work is monotone-increasing with $\beta$ (decreasing with temperature) due to the change of sign of $\frac{\partial^2 F}{\partial\beta\partial\nu}$. In that case, positive net work extraction is achieved when bosonization (work extraction) takes place at the colder temperature, and fermionization happens at the hotter. If $\nu$ crosses the threshold value $\frac{1}{2}N(N{-}1)\hbar\omega$ during the cycle, then the sign of $\frac{\partial^2 F}{\partial\beta\partial\nu}$ is not constant, and the above reasoning cannot be used to determine the sign of work extraction.

\subsection{Stirling cycle: recovering Carnot efficiency}
We here consider the efficiency of the Carnot cycle in the limit as $\nu_1\to -\infty$ and $\nu_2\to +\infty$. Taking an informal approach, from Eq. \eqref{eqn:pf} it can be seen that $\lim_{\nu_2\to +\infty} p_\mathrm{F}(\beta,\nu_2) = 1$ 
Considering the total work for the cycle from Eq. \eqref{cyc_work} as $\nu_2\to +\infty$, we have the following limit for $W_\mathrm{cyc}$
\begin{equation}
    \begin{split}
        W_\mathrm{cyc} &\to \frac{1}{\beta_\mathrm{H}}\ln\left[\frac{{p_\mathrm{F}(\beta_\mathrm{H},\nu_1)}}{1}\right] +  \frac{1}{\beta_\mathrm{C}}\ln\left[\frac{1}{p_\mathrm{F}(\beta_\mathrm{C},\nu_1)}\right].\\
    \end{split}
\end{equation}
On the other hand, as $\nu_1\to -\infty$,
\begin{equation}
    p_\mathrm{F}(\beta,\nu_1) \to \frac{1}{\exp[ \frac{1}{2}N(N{-}1)\beta\hbar\omega - \beta\nu_1  - h(d,N)]},
\end{equation}
so that
\begin{widetext}
\begin{equation}
    \begin{split}
        W_\mathrm{cyc} \to &-\frac{1}{\beta_\mathrm{H}}\ln\left\{\exp[ \frac{1}{2}N(N{-}1)\beta_\mathrm{H}\hbar\omega - \beta_\mathrm{H}\nu_1  - h(d,N)]\right\}+  \frac{1}{\beta_\mathrm{C}}\ln\left\{\exp[ \frac{1}{2}N(N{-}1)\beta_\mathrm{C}\hbar\omega - \beta_\mathrm{C}\nu_1  - h(d,N)]\right\}\\
        &=  -\left[\frac{1}{2}N(N{-}1)\hbar\omega - \nu_1  - \frac{1}{\beta_\mathrm{H}} h(d,N)\right] + \left[\frac{1}{2}N(N{-}1)\hbar\omega - \nu_1  - \frac{1}{\beta_\mathrm{C}} h(d,N)\right]\\
        &= \left(\frac{1}{\beta_\mathrm{H}} - \frac{1}{\beta_\mathrm{C}}\right) h(d,N).
    \end{split}
\end{equation}
\end{widetext}

This gives us the first line of Eq. \eqref{limitingwork}. To find the limit of Eq. \eqref{Qh}, first note that we have already seen that
\begin{equation}\label{d24}
\begin{split}
    &\frac{1}{\beta_\mathrm{H}}\ln\left[\frac{p_\mathrm{F}(\beta_\mathrm{H},\nu_1)}{p_\mathrm{F}(\beta_\mathrm{H},\nu_2)}\right] \to\\
        &\hspace{1.4em}  -\left[ \frac{1}{2}N(N{-}1)\hbar\omega - \nu_1  - \frac{1}{\beta_\mathrm{H}} h(d,N) \right].
    \end{split}
\end{equation}
It remains to find an expression for the difference in internal energy. Since $p_\mathrm{F}(\beta,\nu_2) \to 1$ and $p_\mathrm{F}(\beta,\nu_1) \to 0$, then using Eq. \eqref{eqn:n_particle_internal_energy} followed by Eq. \eqref{eqn:fermion_bosons_internal_energy},
we can write
\begin{equation}\label{d25}
    \begin{split}
        &U(\beta_\mathrm{H},\nu_2) - U(\beta_\mathrm{C},\nu_1) \to U_\mathrm{F}(\beta_\mathrm{H}) - \left[\nu_1 + U_\mathrm{B}(\beta_C)\right]\\
        &= \frac{1}{2}N(N{-}1)\hbar\omega - \nu_1  + \sum_{k=1}^N \frac{k\hbar\omega}{e^{k\beta_\mathrm{H}\hbar\omega}-1} - \frac{k\hbar\omega}{e^{k\beta_\mathrm{C}\hbar\omega}-1}.
    \end{split}
\end{equation}
Combining Eqs. \eqref{d24} and \eqref{d25}, according to Eq. \eqref{Qh}, the heat absorbed is simply
\begin{equation}
    Q_\mathrm{H} \to \frac{1}{\beta_\mathrm{H}} h(d,N) + \sum_{k=1}^N \left( \frac{k\hbar\omega}{e^{k\beta_\mathrm{H}\hbar\omega}-1} - \frac{k\hbar\omega}{e^{k\beta_\mathrm{C}\hbar\omega}-1} \right).
\end{equation}

\section{Details of the fast-switching Otto cycle}\label{app:engine2}

\changes{The internal energy of the system following each of the strokes is summarized in the table below:

\setlength{\tabcolsep}{5pt}
\renewcommand{\arraystretch}{1.5}
\begin{center}
\begin{tabular}{ | m{2.5cm} | m{3cm}| m{1.5cm} | } 
  \hline
  \textbf{Stroke} & \textbf{Internal energy following stroke} & \textbf{Transfer type} \\ 
  \hline
  \textbf{1.} Heat & $U_1 = U(\beta_\mathrm{H},\omega_1)$ & $Q_\mathrm{H}$ \\ 
  \textbf{2.} Expand & $U_2 = \frac{\omega_2}{\omega_1}U(\beta_\mathrm{H},\omega_1)$ & $W_\mathrm{out}$ \\
  \textbf{3.} Cool & $U_3 = U(\beta_\mathrm{C},\omega_2)$ & $Q_\mathrm{C}$ \\
  \textbf{4.} Compress & $U_4 = \frac{\omega_1}{\omega_2} U(\beta_\mathrm{C},\omega_2)$ & $W_\mathrm{in}$ \\
  \hline
\end{tabular}
\end{center}

\hfill

From the above table it should be clear that the heat absorbed from the hot reservoir is
\begin{equation}
    \begin{split}
        Q_\mathrm{H} &= U_1 - U_4\\
        &= U(\beta_\mathrm{H},\omega_1) - \frac{\omega_1}{\omega_2} U(\beta_\mathrm{C},\omega_2).
    \end{split}
\end{equation}
Then, noting from Eq. \eqref{eqn:fermion_bosons_internal_energy} that $U(\beta,\omega) = U_\mathrm{B}(\beta,\omega) + \frac{1}{2}N(N-1)\hbar\omega\, \pf(\beta,\omega)$, we have
\begin{equation}\label{ottoheat}
    \begin{split}
        Q_\mathrm{H} &= U_\mathrm{B}(\beta_\mathrm{H},\omega_1) + \frac{1}{2}N(N-1)\hbar\omega_1\, \pf(\beta_\mathrm{H},\omega_1) \\ 
        &\hspace{1em}- \frac{\omega_1}{\omega_2} \left[ U_\mathrm{B}(\beta_\mathrm{C},\omega_2) - \frac{1}{2}N(N-1)\hbar\omega_2\, \pf(\beta_\mathrm{C},\omega_2)\right]\\
        &= U_\mathrm{B}(\beta_\mathrm{H},\omega_1)  - \frac{\omega_1}{\omega_2}  U_\mathrm{B}(\beta_\mathrm{C},\omega_2)\\
        & \hspace{1em} + \frac{1}{2} N(N-1)\hbar\omega_1\, \Big[ \pf(\beta_\mathrm{H},\omega_1) - \pf(\beta_\mathrm{C},\omega_2)\Big].
    \end{split}
\end{equation}
Note in particular that the last line of the above vanishes whenever $\pf$ is fixed: that is, the same amount of heat is absorbed by pure bosons and pure fermions, with a different amount absorbed by Hamiltonian anyons. Something similar happens for the net work output of the cycle:
\begin{equation}\label{ottowork}
    \begin{split}
        W_\mathrm{cyc} &= (U_1 - U_2) + (U_3 - U_4)\\
        &= \left(1 - \frac{\omega_2}{\omega_1}\right) U(\beta_\mathrm{H},\omega_1) + \left(1 - \frac{\omega_1}{\omega_2}\right) U(\beta_\mathrm{C},\omega_2)\\
        &= \left(1 - \frac{\omega_2}{\omega_1}\right) U_\mathrm{B}(\beta_\mathrm{H},\omega_1) + \left(1 - \frac{\omega_1}{\omega_2}\right) U_\mathrm{B}(\beta_\mathrm{C},\omega_2)\\
        &\hspace{1em} + \frac{1}{2}N(N{-}1)\hbar(\omega_1 {-} \omega_2)\Big[\pf(\beta_\mathrm{H},\omega_1) - \pf(\beta_\mathrm{C},\omega_2)\Big].\\
    \end{split}
\end{equation}
Here again, the final line is only nonzero where $\pf$ is variable, meaning that $W_\mathrm{cyc}$ is the same for pure bosons and fermions. For a working medium of Hamiltonian anyons driven across the transition such that $\pf(\beta_\mathrm{H},\omega_1) - \pf(\beta_\mathrm{C},\omega_2) \approx 1$, the excess work per cycle scales as $N^2$.

With a little rearranging, we can also find a simple expression for the efficiency $\eta = \frac{W_\mathrm{cyc}}{Q_\mathrm{H}}$. In particular, we can factor out $(\omega_1 - \omega_2)$ from Eq. \eqref{ottowork} as
\begin{equation}
    \begin{split}
        W_\mathrm{cyc} &= (\omega_1 - \omega_2)\Bigg[ \frac{1}{\omega_1} U_\mathrm{B}(\beta_\mathrm{H},\omega_1) - \frac{1}{\omega_2} U_\mathrm{B}(\beta_\mathrm{C},\omega_2) \\
        &\hspace{3em} + \frac{\hbar}{2}N(N{-}1)\Big(\pf(\beta_\mathrm{H},\omega_1) - \pf(\beta_\mathrm{C},\omega_2)\Big)\Bigg].\\
    \end{split}
\end{equation}
Likewise, $\omega_1$ can be pulled out of Eq. \eqref{ottoheat} leaving the same factor in square brackets:
\begin{equation}\label{ottoheat2}
    \begin{split}
        Q_\mathrm{H} &= \omega_1 \Bigg[ \frac{1}{\omega_1} U_\mathrm{B}(\beta_\mathrm{H},\omega_1) - \frac{1}{\omega_2} U_\mathrm{B}(\beta_\mathrm{C},\omega_2) \\
        &\hspace{3em} + \frac{\hbar}{2}N(N{-}1)\Big(\pf(\beta_\mathrm{H},\omega_1) - \pf(\beta_\mathrm{C},\omega_2)\Big)\Bigg].\\
    \end{split}
\end{equation}
Thus, it becomes clear that the efficiency for fermions, bosons and Hamiltonian anyons is
\begin{equation}\label{ottoeta}
    \eta = 1 - \frac{\omega_2}{\omega_1},
\end{equation}
uniquely specified by the compression ratio. Up to now, we have tacitly assumed that $Q_\mathrm{H} = Q_\mathrm{in}$:  that a positive amount of heat is absorbed from the hot bath and positive heat is expelled to the cold bath (in other words, the cycle operates as an engine). Actually, we can check 
under what conditions this assumption holds. From Eq. \eqref{eqn:fermion_bosons_internal_energy} it is straightforward to see that $\frac{1}{\beta} U(\beta,\omega)$ is a monotone-decreasing function of the product $\beta\omega$. Since we fixed $\nu=0$}\changes{, it is also clear from Eq. \eqref{eqn:pf} that $\pf$ is monotone-decreasing in $\beta\omega$. Therefore the square-bracketed term appearing in both Eqs. (\ref{ottowork})--(\ref{ottoheat}) is positive if and only if $\beta_\mathrm{H}\omega_1<\beta_\mathrm{C}\omega_2$. As a consequence, we recover the expected result that the otto cycle efficiency Eq. \eqref{ottoeta} cannot exceed the Carnot efficiency in regimes where the cycle operates as an engine. Moreover, we see that in the limit as $\beta_\mathrm{H}\omega_1$ tends to $\beta_\mathrm{C}\omega_2$ from above, the cycle can approach Carnot efficiency but at the expense of vanishing work output.}

\end{document}